\PassOptionsToPackage{x11names,svgnames}{xcolor}
\documentclass[aps,preprint]{revtex4}
\usepackage{amssymb}
\usepackage{amsmath}
\usepackage{xspace}
\usepackage{graphicx}
\usepackage[export]{adjustbox}
\usepackage{subfig}
\usepackage{placeins}

\usepackage{siunitx}
\DeclareSIUnit{\electronvolt}{\text{e\kern-0.15ex V}}
\DeclareSIUnit{\eV}{\electronvolt}
\DeclareSIUnit{\MeV}{\mega\eV}
\DeclareSIUnit{\GeV}{\giga\eV}
\DeclareSIUnit{\TeV}{\tera\kern-0.1ex\eV}

\usepackage{mydefs}

\usepackage{tikz}

\newcommand{\swatch}[1]{\tikz[baseline=-0.6ex]\node[fill=#1,shape=rectangle,draw=black,thick,minimum width=5mm,rounded corners=0.5pt](){};}
\newcommand{\met}{\ensuremath{E_T^{\rm miss}}}

\definecolor{deepskyblue}{HTML}{00BFFF}
\definecolor{green}{HTML}{008000}
\definecolor{darkolivegreen}{HTML}{556B2F}
\definecolor{orangered}{HTML}{FF4500}
\definecolor{crimson}{HTML}{DC143C}
\definecolor{seagreen}{HTML}{2E8B57}
\definecolor{cornflowerblue}{HTML}{6495ED}
\definecolor{silver}{HTML}{C0C0C0}
\definecolor{blue}{HTML}{0000FF}
\definecolor{dimgrey}{HTML}{696969}
\definecolor{darkorange}{HTML}{FF8C00}
\definecolor{darksalmon}{HTML}{E9967A}
\definecolor{snow}{HTML}{FFFAFA}
\definecolor{navy}{HTML}{000080}
\definecolor{turquoise}{HTML}{30D5C8}
\definecolor{powderblue}{HTML}{B0E0E6}
\definecolor{yellow}{HTML}{FFFF00}
\definecolor{orange}{HTML}{FFA500}
\definecolor{lightsalmon}{HTML}{FFA07A}
\definecolor{tomato}{HTML}{FF6347}
\definecolor{mediumseagreen}{HTML}{3CB371}
\definecolor{magenta}{HTML}{FF00FF}
\definecolor{darkgoldenrod}{HTML}{B8860B}
\definecolor{cadetblue}{HTML}{5F9EA0}
\begin{document}

\title{
New sensitivity of LHC measurements to Composite Dark Matter Models
}

\author{J. M. Butterworth}
\email{J.Butterworth@ucl.ac.uk}
\author{X. Kong}
\author{M. Thomas}
\affiliation{Department of Physics \& Astronomy, UCL, Gower~St., WC1E~6BT, London, UK}

\author{L. Corpe}
\email{l.corpe@cern.ch}
\affiliation{CERN, Esplanade des Particules 1, 1211 Geneva, Switzerland}

\author{S. Kulkarni}
\email{suchita.kulkarni@uni-graz.at}
\affiliation{Institute of Physics, NAWI Graz, University of Graz, Universit\"atsplatz 5, A-8010 Graz, Austria}

\date{\today}


\begin{abstract}
We present sensitivity of LHC differential cross-section measurements to so-called ``stealth dark matter'' 
scenarios occurring in an \sund dark gauge group, where constituents are charged under the Standard Model and \nd = 2 or 4. 
The low-energy theory contains mesons which can be produced at the LHC, and a scalar baryon dark matter (DM) candidate which cannot. 
We evaluate the impact of LHC measurements on the dark meson masses. Using existing lattice results, we then connect the LHC 
explorations to DM phenomenology, in particular considering direct-detection experiments. 
We show that current LHC measurements constrain DM masses in the region of 3.0-5.5~TeV. 
We discuss potential pathways to explore these models further at the LHC.
\end{abstract}


\maketitle

\section{Introduction}
\label{sec:intro}

Strongly-interacting theories featuring new non-Abelian gauge groups, where confinement in a ``dark sector'' (DS) at some confinement scale $\Lambda$ leads to stable composite states, offer an interesting alternative explanation of Dark Matter (DM). 
In these theories, the stability of the DM candidate can be ensured either by imposing additional symmetries or, more naturally, result from accidental symmetries of the theory. The dark matter candidates can thus be dark pions, baryons or glueballs depending on the exact setup.
Such theories can be realised in a variety of non-Abelian gauge groups and can even help explain observed discrepancies between observation and simulations at cosmic scales via so-called DM self-interactions~\cite{Hochberg:2014dra,Tsai:2020vpi,Lee:2015gsa,Choi:2016hid,Cline:2013zca,Boddy:2014yra,Boddy:2014qxa,Soni:2016gzf}. Moreover, in addition to DM candidate(s), much like Standard Model (SM) QCD such dark non-Abelian gauge theories feature a spectra of bound states. While such composite DM candidates may exist as a result of strong dynamics, whether and how these DS theories communicate with the SM remains an interesting open question. To this end, one could introduce new SM--DS mediators~\cite{Beauchesne:2018myj,Hochberg:2018rjs,Bernreuther:2019pfb,Mies:2020mzw,Renner:2018fhh}, or charge DS fermions under some of the SM gauge group~\cite{Buckley:2012ky,Appelquist:2015yfa,Bai:2010qg,Kribs:2009fy,Appelquist:2013ms}. The latter scenarios were also realised in the context of technicolor theories e.g.~\cite{Nussinov:1985xr, 1990NuPhB.329..445C,Barr:1990ca,Ryttov:2008xe,Lewis:2011zb}, at least some of which are now under siege after the discovery of the SM Higgs boson; when the DS fermions carry a SM charge, careful consideration of existing electroweak precision tests is required. For a review on fundamental composite dynamics and discussions of SM electroweak constraints see~\cite{Cacciapaglia:2020kgq}.

Once a non-Abelian gauge group with fixed number of flavours and colours is chosen, together with a fermionic representation and a SM--DS mediation mechanism, theoretical predictions for the mass spectra of bound states are obtained by means of lattice simulations. These simulations can predict several useful quantities for a phenomenological analysis in three different regimes; 
(i) the chiral regime where dark quarks can be assumed massless $m_{q_D} \ll \Lambda$, (ii) the comparable scales regime $m_{q_D} \sim \Lambda$, and 
(iii) the heavy quark/quarkonia regime $m_{q_D} \gg \Lambda$. 
Along with the dark hadron mass spectra, lattice simulations can also predict a variety of useful inputs, such as decay constants or matrix elements, 
useful for computing cross-sections. These predictions are then taken as an input for a low-energy effective theory in order to devise experimental searches and evaluate sensitivities. It should be noted that lattice simulations do not predict exact mass-scales. However they can predict bound state masses in terms of some common mass scale, which may be freely chosen. For an excellent review pertaining to this discussion see~\cite{Kribs:2016cew}.

Experimental searches for such strongly-interacting dark sectors depend on the mediator mechanisms as well as the comparative mass scales. For example, at the LHC, cases where dark quark masses ($m_{q_D}$) and corresponding confinement scale are much smaller than the collider centre-of-mass energy ($m_{q_D} \lesssim \Lambda \ll \sqrt{s}$) lead to spectacular signatures in terms of semi-visible jets or emerging jets~\cite{Cohen:2015toa,Daci:2015hca,Schwaller:2015gea}. If on the other hand if the three scales are comparable ($m_{q_D} \sim \Lambda \sim \sqrt{s}$), then resonance-like searches may prove useful, depending on the relevant production mechanisms~\cite{Kribs:2018ilo,Hochberg:2015vrg,Kribs:2016cew}. Finally, cases where $m_{q_D} \gg \Lambda,  m_{q_D} \lesssim \sqrt{s}$ lead to unusual signals known as quirks~\cite{Knapen:2016hky, Harnik:2008ax}. If the strongly-interacting sector is non-QCD like, other signatures such as Soft Unclustered Energy Patterns are also possible~\cite{Knapen:2016hky, Harnik:2008ax}. 

In the vast program of exploring strongly-interacting theories, direct searches for such scenarios have been a focus of the experimental program~\cite{Sirunyan:2018njd, CMS:2020kwj}. 
In this work, we instead demonstrate the power of precision measurements of SM-like final states, by taking the so-called ``stealth dark matter" scenarios~\cite{Kribs:2018ilo} as an example theory. Stealth dark matter scenarios are realised in \sund theories with even \nd. In such theories, the baryonic DM candidate is a scalar particle and is stable on account of dark baryon number conservation.
Along with the dark baryon, the theory also features dark pions and mesons as bound states, which are lighter than the dark baryon.

Dark sector interactions with the SM are realised by charging part (or all) of the dark sector under the SM electroweak gauge group. 
This leads to signals at direct-detection experiments via Higgs exchange, and the dark rho (\rhod) mixing with the SM gauge bosons leads to signals at the LHC. 
Kribs et al~\cite{Kribs:2018ilo} considered such a theory and performed generic lattice simulations for $\nd = 4$ in the comparable scales regime 
$m_q \sim \Lambda$ and the quenched limit. They furthermore constructed concrete realisations of such a model where dark quarks respect exact custodial SU(2) symmetry~\cite{Appelquist:2015yfa}. In  \cite{Kribs:2018oad} they constructed effective theories for dark mesons in such theories while in \cite{Kribs:2018ilo}, they confronted the meson sector with LHC searches. In doing so, they have provided a complete setup from microscopic theory of dark quarks to macroscopic theory of scalar dark matter and meson bound states. With a mass scale ranging from $\mathcal{O}(100) \rm{GeV}$ to TeV, this theory is an ideal candidate with which to explore the implications of LHC cross-section measurements for strongly-interacting DM scenarios. In this paper we use \CONTUR~\cite{Butterworth:2016sqg,Buckley:2021neu} to study the impact such bound states would have had on existing LHC measurements, and using the lattice calculations of Appelquist et al~\cite{Appelquist:2014jch,Appelquist:2015yfa}, connect this to the relevant cosmological and direct-detection limits.
The \CONTUR method makes use of the bank of LHC measurements (and a few searches) whose results and selection logic are preserved in runnable \RIVET~\cite{Bierlich:2019rhm} routines.
Hundreds of measurements are preserved in this way, which allows generated signal events to quickly and efficiently be confronted with the observed data in a wide variety of final states.
In simple terms, if a new-physics contribution would have modified a SM spectrum beyond its measured uncertainties, ``we would have seen it''; \CONTUR uses the CL$_{s}$~\cite{Read:2002hq} method to quantify this exclusion, making use of bin-to-bin uncertainty correlation information from LHC measurements where available.
This approach has been shown to be highly complementary to the direct search programme~\cite{Buckley:2020wzk,Butterworth:2020vnb}. For models such as those discussed in this paper, where the expected signature at a $pp$ collider changes drastically depending on the model parameter choices, direct searches may be inefficient and hard to motivate. \CONTUR offers a comprehensive and robust way to probe the parameter space.

The paper is structured as follows. In the next section the models  are summarised, while Section~\ref{sec:collider} is dedicated to the resulting collider phenomenology. In Section~\ref{sec:contur} we
present and discuss the implications of LHC measurements for the putative dark mesons. In Section~\ref{sec:dm} we then translate these constraints
into constraints on the DM candidate and discuss the impact on DM phenomenology more generally, before concluding.

\section{Model details}
\label{sec:model}
The principal model considered here is a \sund gauge theory with $\nd = 4$ and $N_f = 4$ (Weyl flavours), and the dark quarks (fermions) are in the fundamental representation of the dark colour gauge group \sund\footnote{We will also briefly discuss the case $\nd=2$}. 
The dark quarks are further charged under the SM gauge group and transform in a vector-like representation. As vector-like fermions, they have a mass term 
which is independent of any electroweak symmetry breaking mechanism. Charging them under the SM gauge group nevertheless generates interactions with 
the SM Higgs boson, and lead to masses originating from electroweak symmetry breaking just like any other SM fermion masses. 
It is possible to write down these renormalisable vector and chiral mass terms in $N_f = 4$ theory, though not in the  $N_f = 2$ theory~\cite{Kribs:2018oad}.  For simplicity in the theory, $m$ characterises a common vector-like mass term with $\Delta$ introducing the splitting, while $yv$ characterises the chiral mass with a small factor $\epsilon$ enabling splitting of the chiral masses. Here we will assume $\epsilon$ is negligible. 
Electroweak precision tests and Higgs coupling measurements constrain the electroweak symmetry breaking mass term to
be small compared to the vector masses ($yv \ll m$).

In the absence of charges under the SM -- in other words when the theory is taken in isolation -- it exhibits 
SU($N_D^{\textrm{fund}}$) $\times$ SU($N_D^{\textrm{anti}}$) chiral symmetry, where $N_D^{\textrm{fund}}$ denotes fundamental representation and $N_D^{\textrm{anti}}$ the anti-fundamental. 
When couplings to the SM Higgs boson are turned on, some of the flavour symmetries are explicitly broken. 
The model under consideration here will however preserve custodial SU(2) which is the residual accidental global symmetry of the Higgs multiplet after it acquires a vacuum expectation value. Interactions with the Higgs connect flavor symmetries of the fermionic sector with the $O(4) \simeq \sutl \times \sutr$ global symmetry Higgs potential. Out of these the \sutr group contains the U(1)$_\textrm{Y}$ subgroup of the SM via the $t_3$ generator of the SU(2) gauge group. 

Below the dark confinement scale, the theory becomes a low-energy effective field theory and is described in terms of mesons and baryons of the sector. 
Confinement spontaneously breaks the chiral symmetry of the dark fermions down to the diagonal subgroup $SU(N_D^{\textrm{fund}}) \times SU(N_D^{\textrm{anti}}) \to \sund_V$, 
where dark pions live. In total there are 15 pions corresponding to $N_f^2 - 1$ broken generators. 
However, for the purposes of the phenomenology, only the lightest pions of the theory, \pid, are considered. 
Similarly there are dark rho mesons, \rhod. The \pid and \rhod form triplets under \sutl or \sutr . Should the \sutl be gauged, the \rhod 
mix with all three SM weak gauge bosons, and can thus be produced at the LHC via the Drell-Yan (DY) process. 
For the \sutr case only the \rhodz
can be produced this way, since when one chooses to gauge the U(1)$_{\textrm{Y}}$ subgroup only the \rhodz mixes,
in this case with the SM $B$ field.

The decays of \pid are also interesting and are intimately connected to the mass spectrum and symmetries of the theory. 
The dark pions (to be precise, dark kaons) of the theory mix with the Goldstones of the SM and thus generate couplings with the SM gauge bosons as well with the Higgs boson. 
It can be shown with the help of chiral perturbation theory that decays of the \pid to gauge bosons are suppressed by
$\sim m_h^2/m^2_{K_D}$, in both \sutl and \sutr scenarios.
(Here $m_{K_D}$ is the mass of dark kaon,
which is assumed to be not much heavier than \mpid, the mass of the \pid.)
The models of greatest interest to our discussion are therefore referred to as `gaugephobic' \sutl and \sutr scenarios. 
Conceptually this small coupling to the gauge bosons can be understood as the Higgs mixing with the kaons, which are doublets,
leading to a suppression by a factor of $m_h^2/(m^2_{K_D}-m_h^2)$, which approximates to $\sim m_h^2/m^2_{K_D}$ when the dark kaon
masses are heavier than the Higgs mass.

\begin{figure}[tbp]
\begin{centering}
\includegraphics[width=0.7\textwidth]{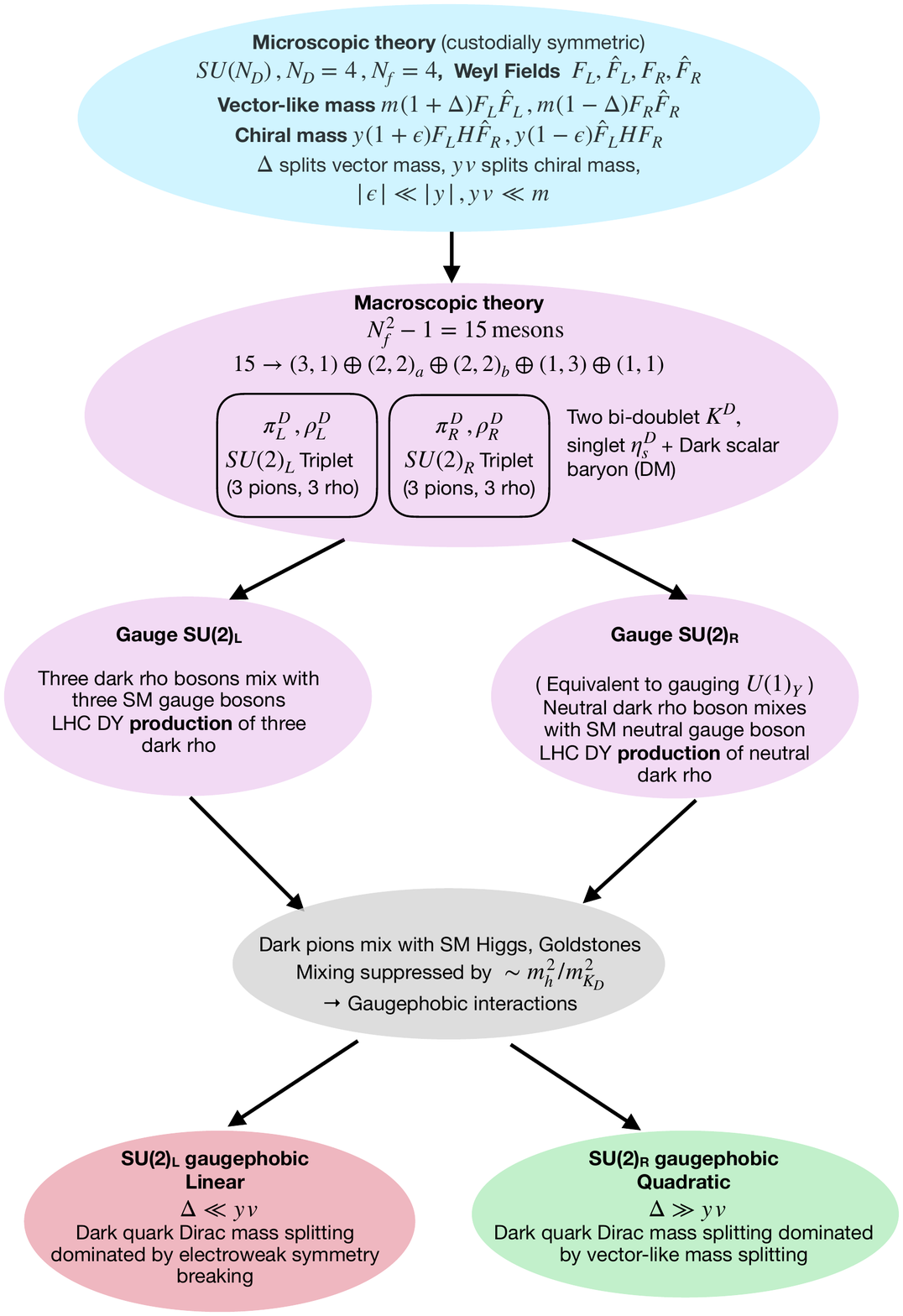}
\caption{Flowchart of the model parameters details. Throughout this work, we will use low-energy parametrization of the theory.  Fields depicted as $\pi^D_L, \rho^D_L, \pi^D_R, \rho^D_R$ are referred to as \pid and \rhod in the text without $L, R$ indices, since they serve no purpose other
than to illustrate multiplicity.  }
\label{fig:model_details}
\end{centering}
\end{figure}

Finally, the origin of DM phenomenology in this model deserves some discussion.
The model features a scalar dark baryon which is stable by virtue of dark baryon symmetry. 
As the dark quark masses have contributions from electroweak symmetry breaking, it also gives rise to scalar baryon (DM) interactions
with the Higgs boson, leading to a Higgs-mediated signal at direct-detection experiments. 
The two mass terms also imply that dark quarks, and consequently the dark baryon, have a tuneable coupling to the Higgs boson. 

Due to the strongly-interacting nature of the theory, lattice computations prove to be useful for determining the inputs for the low-energy effective 
theory. For this model, such lattice simulations were performed for quenched masses in comparable regime scenario $\Lambda \sim m_{q_D}$. 
These simulations provide us with the spectra of \pid, \rhod and dark baryon masses for certain values of the pion to rho mass ratio 
($\mpid/\mrhod \equiv \eta$) in units of a dimensionful scaling parameter. 
Along with this, the calculations provide the matrix element for dark baryon scattering via Higgs at direct-detection experiments, parametrised 
as $f^{DM}_f$ for the same values of $\eta$. We use these quantities as inputs for our study. 
We will ultimately derive constraints the tuneable Higgs to dark quark coupling. 

The above discussion is summarised as a flowchart in Figure~\ref{fig:model_details}. For the sake of clarity we use explicit left and right indices for $\pi_D, \rho_D$ in the figure. In the discussions however, we drop these indices assuming that the dark rhos and pions belong to the representation of choice. 

In practical terms, we use the effective Lagrangians as implemented in \cite{Kribs:2018ilo}\footnote{Except for changing $N_D$, $\eta$, \mpid we keep all other inputs in the model files to be default.}. This Lagrangian does not contain dark kaon production, despite them being at a similar mass scale to the dark pion. The dark kaons are expected to be produced via their mixing with the SM Higgs bosons and will in general lead to small a production cross section, due to the mixing suppression and high masses. Thus, only phenomenology of dark pions and rhos is considered.

For the low-energy parametrization, one can write an effective Lagrangians where dark pion couplings to gauge bosons are suppressed by $\xi \sim m_h^2/m^2_{K_D}$, as discussed above, or take $\xi$ to be a free parameter and set it to 1. When $\xi = 1$, the pion couplings to gauge bosons are unsuppressed and this leads to a so-called `gaugephilic' scenario. For such a scenario to exist, the dark pions should be in the $SU(2)_L$ representation and hence (only) the low-energy parametrization of the $SU(2)_L$ model involves a gaugephilic scenario as well. 

An example of such a gaugephilic scenario can be realised in, for example, two-flavour chiral theory, and is discussed at length in \cite{Kribs:2018oad}. For theories containing two flavours, there is no scalar baryon hence no dark matter candidate.
We will discuss gaugephilic scenarios from the point of view of their collider phenomenology; however no connection to the DM phenomenology
will be made.

\section{Dark mesons and collider phenomenology}
\label{sec:collider}

As discussed above, the kinematic mixing of the \rhod with SM gauge bosons opens the possibility of its production via quark-antiquark annihilation (DY).
If kinematically allowed, the decay of the \rhod to \pid pairs is possible. The \rhod may also decay directly to SM fermion-antifermion pairs.
In the gaugephobic case, the \pid decays primarily to pairs of fermions, while in gaugephilic cases it may also decay via $W+h$ or $Z+h$ 
if this is kinematically allowed. 

\subsection{\rhod and \pid production and decay at the LHC}

In our study we calculate the cross-section for \rhod and \pid production using \HERWIG~\cite{Bellm:2019zci} to simulate events based upon the
Feynrules model files~\cite{Degrande:2011ua} provided by Kribs et al~\cite{Kribs:2018ilo}. We scan over various parameter planes of the model, 
generating all leading-order \TwoToTwo processes in which at least one dark meson is either an outgoing leg, 
or an $s$-channel resonance.

The most important dark meson production mechanism at the LHC is single \rhod production, often in association with another hard particle, 
with the cross-sections for \sutl models typically an order of magnitude larger than for \sutr models. 

Assuming 13~TeV $pp$ collisions, the largest cross-sections come from \rhod particles produced with a quark or gluon 
(for a $\mrhod \sim 1$~TeV, of the order of 100~fb each for \rhodz, 
\rhodp and \rhodm for \sutl, 
and of the order of 10~fb, for \rhodz only, in \sutr).
The \rhod can also be produced in association with a weak vector boson, with a cross-section about an order of magnitude less than for quarks and 
gluons in the \sutl case. In the \sutr case, production with weak bosons is negligible.
Finally, the \rhod can be produced with a photon, with cross-sections around $1$ and $0.1$~fb for \sutl and \sutr respectively.
\HERWIG will also generate \pid pair production mediated via an $s$-channel \rhod, taking into account the calculated \rhod 
width\footnote{There is an overlap between these processes and the $\rhod$+jet processes, which are divergent at low jet transverse momentum.
\HERWIG regulates this by imposing a cut-off on the transverse momentum of the outgoing partons which defaults to 20~GeV. 
The total visible cross-section, and the sensitivities derived in the next section, have a very weak dependence on this cut-off for 
values of a few tens of GeV.}.

\begin{figure}[tbp]
\vspace{-0.4cm}
\subfloat[$pp \rightarrow \rhodz q$ (\sutl)]{\includegraphics[width=0.40\textwidth]{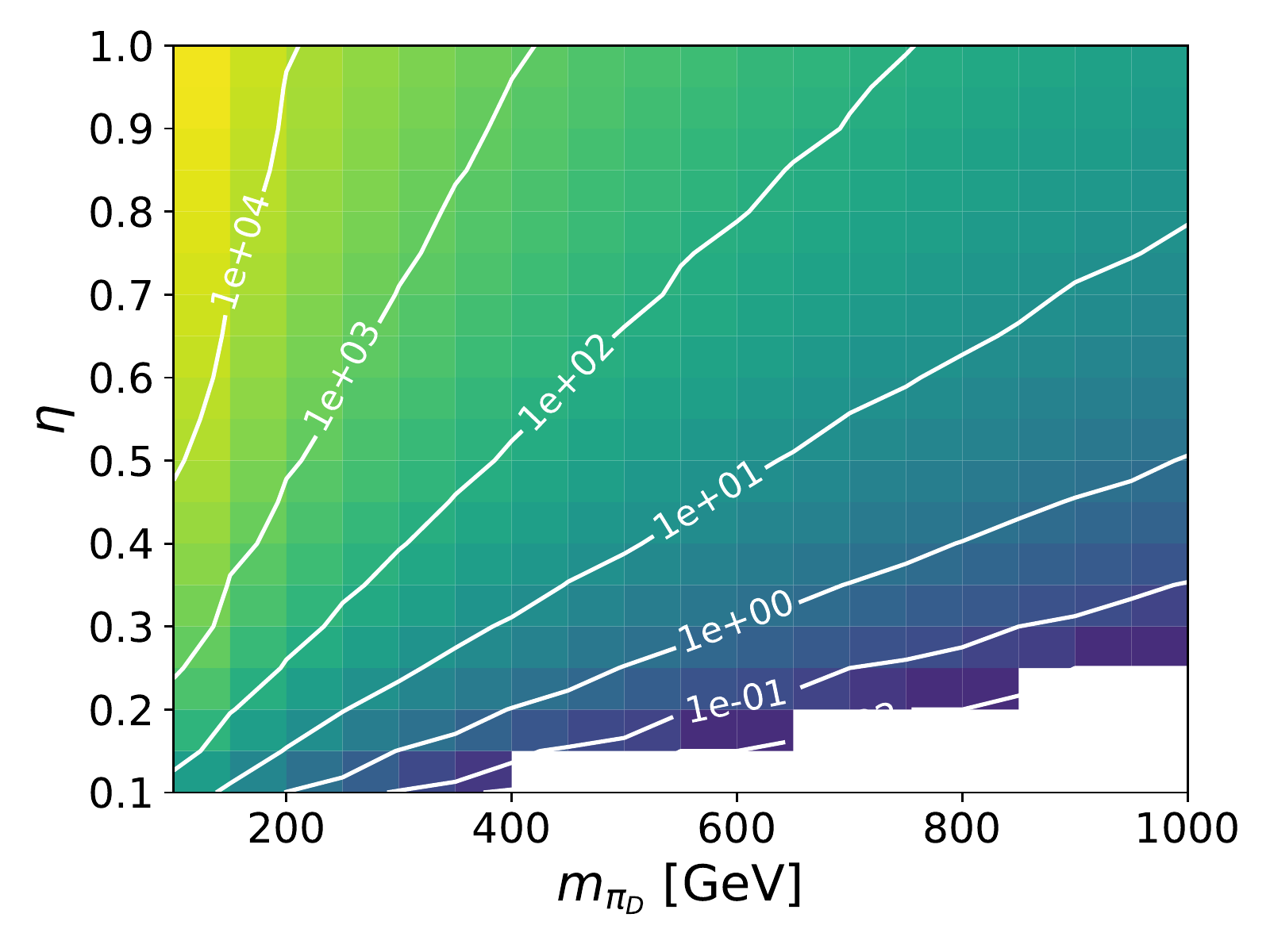}}
\subfloat[$pp \rightarrow \rhodz q$ (\sutr)]{\includegraphics[width=0.40\textwidth]{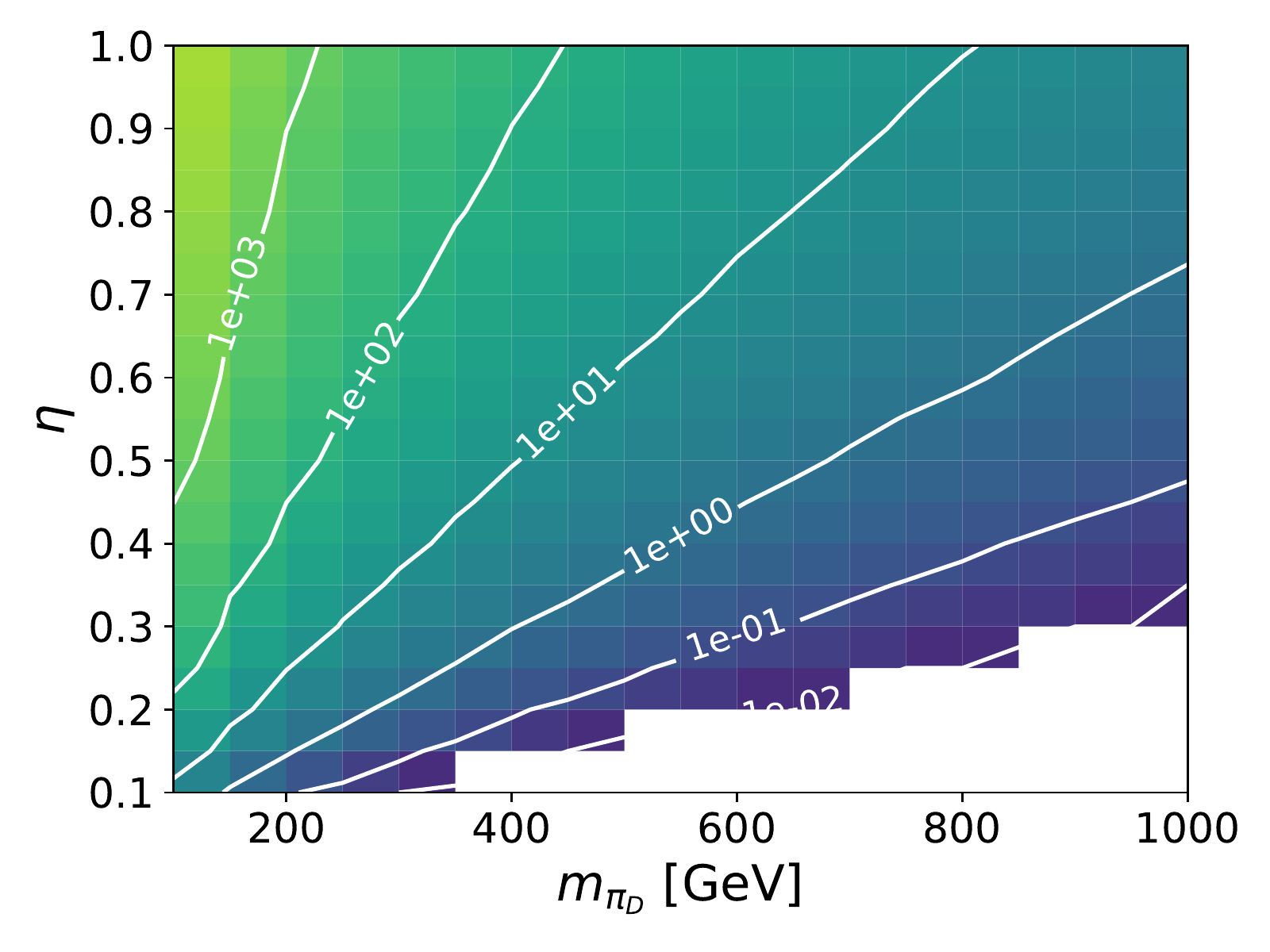}} 
\includegraphics[height=4.5cm]{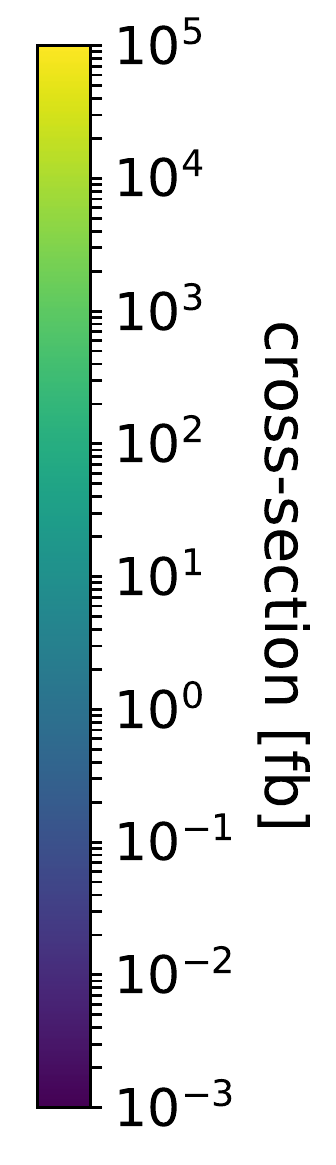} \\ 
\subfloat[$pp \rightarrow \pidp \pidm$ (\sutl)]{\includegraphics[width=0.40\textwidth]{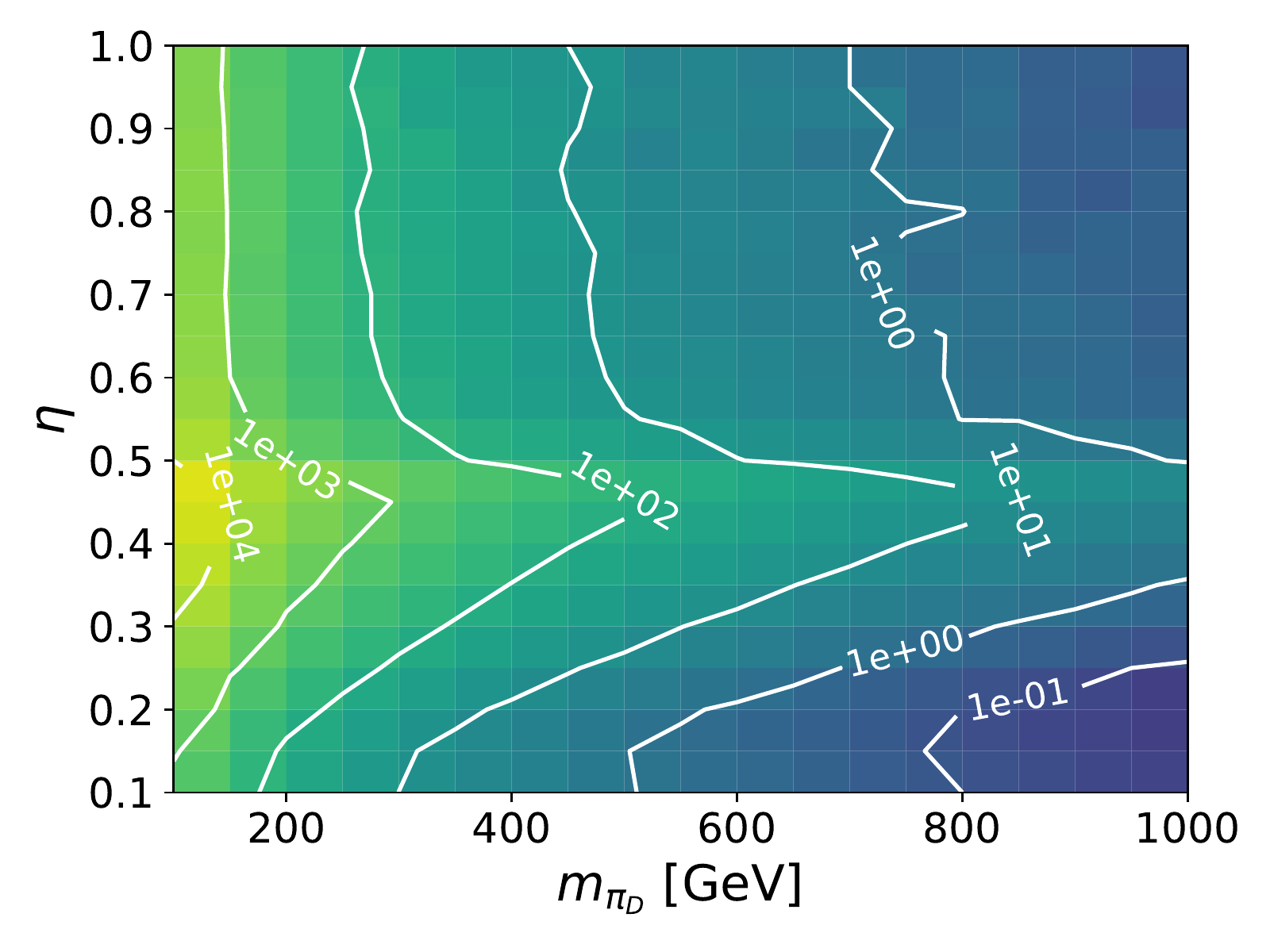}}
\subfloat[$pp \rightarrow \pidp \pidm$ (\sutr)]{\includegraphics[width=0.40\textwidth]{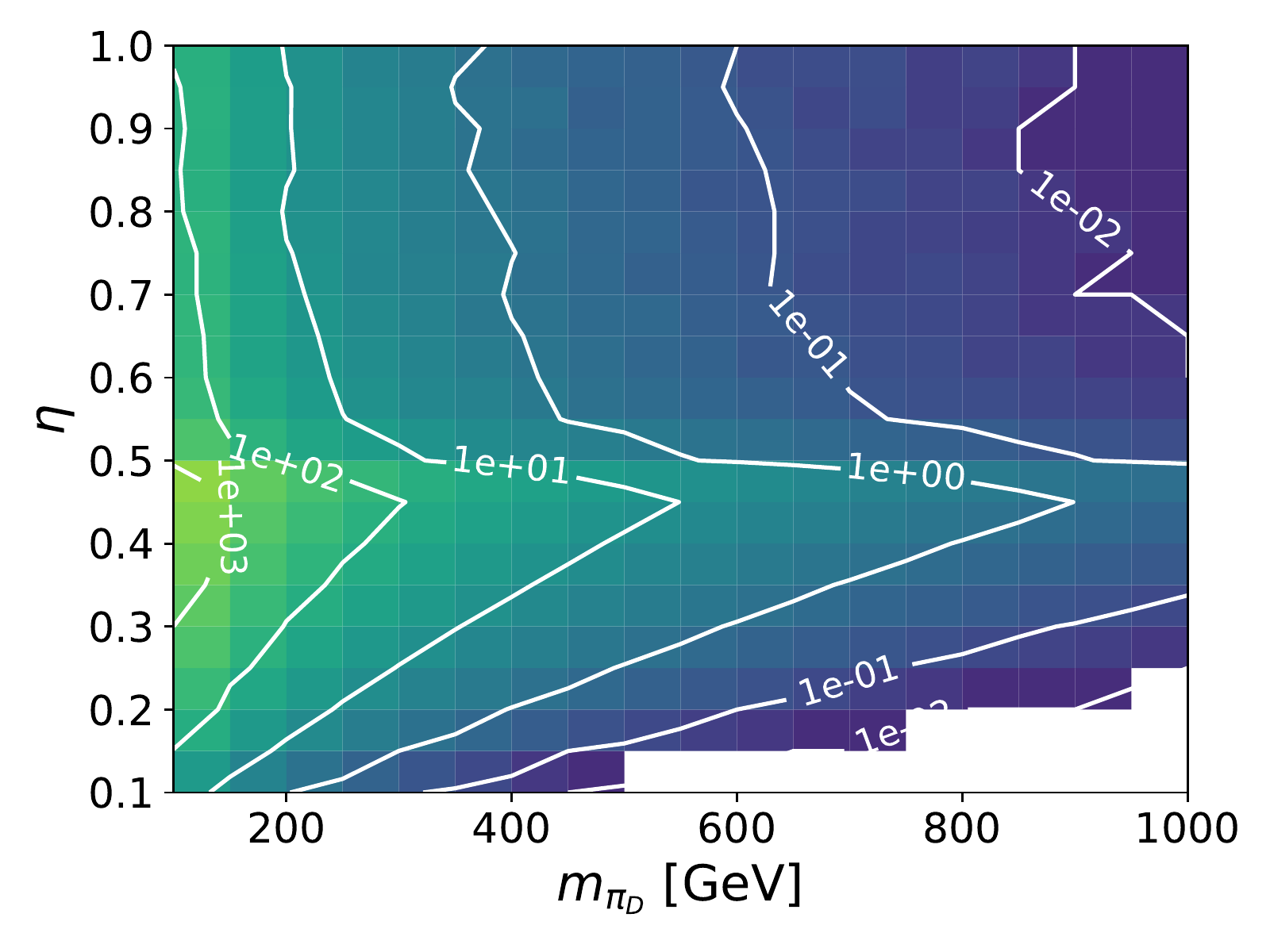}} 
\includegraphics[height=4.5cm]{figures/cross_sections/GoL_v4/cbar.pdf} \\ 
\subfloat[$s$-channel $pp\rightarrow \rhod \rightarrow l^+ l^- $ (\sutl)]{\includegraphics[width=0.40\textwidth]{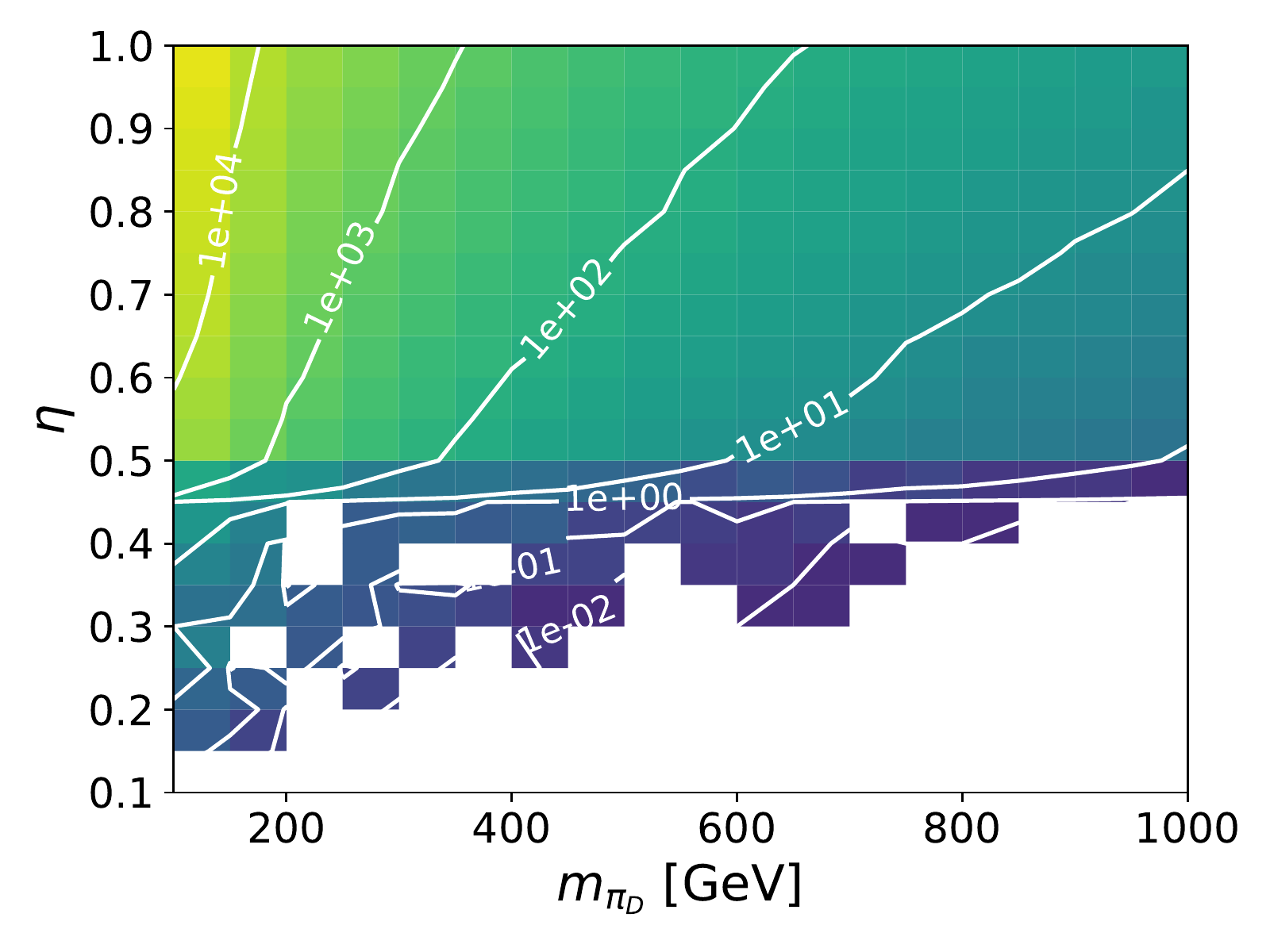}} 
\subfloat[$s$-channel $pp\rightarrow \rhod \rightarrow l^+ l^- $ (\sutr)]{\includegraphics[width=0.40\textwidth]{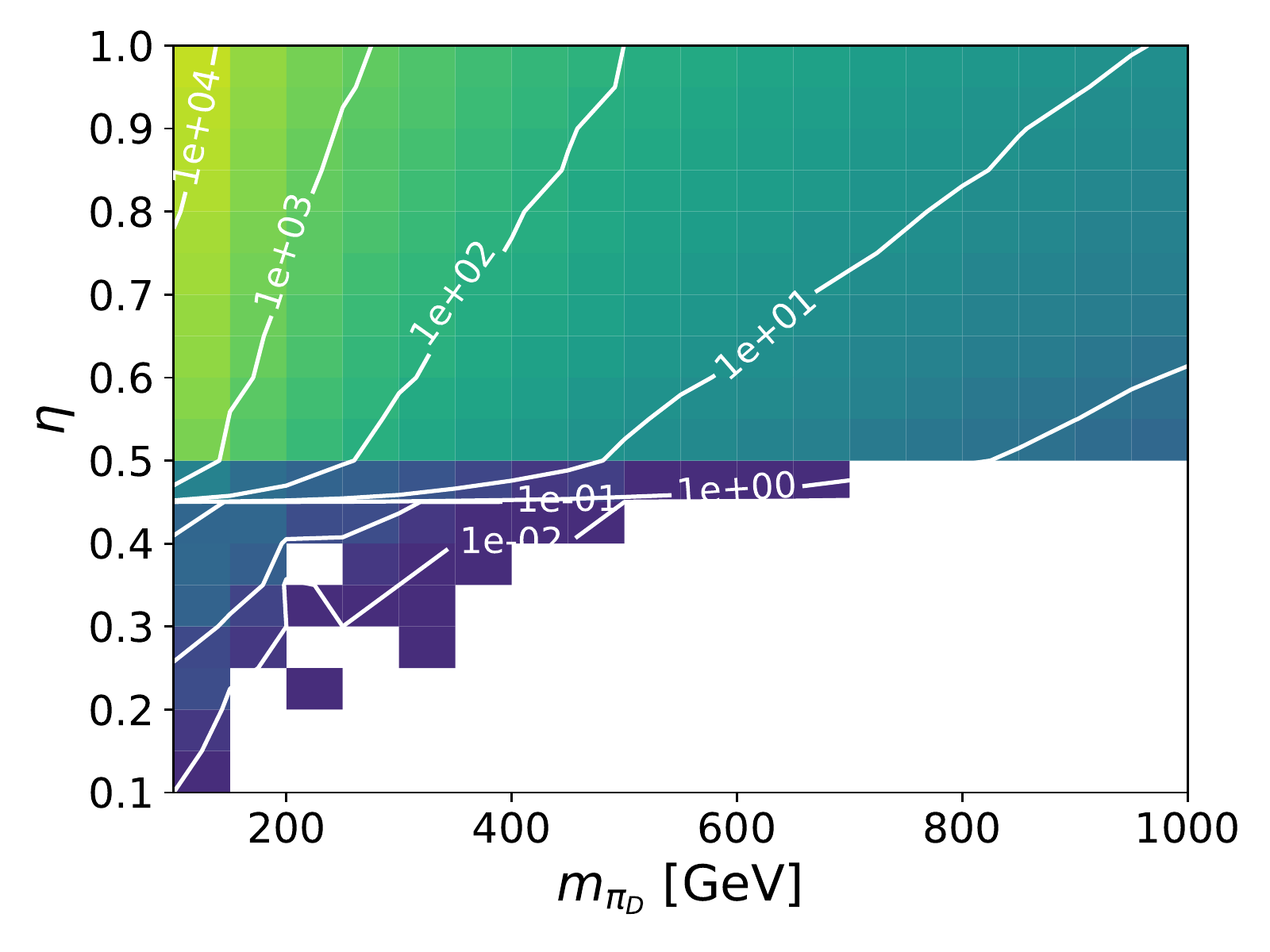}} 
\includegraphics[height=4.5cm]{figures/cross_sections/GoL_v4/cbar.pdf} \\ 
\caption{Selected BSM cross-sections as a function of \mpid and $\eta$, for 13~TeV $pp$ collisions. 
The left (a,c,e) and right (b,d,f) columns shows cross-sections for \sutl and \sutr respectively. The shapes are qualitatively similar, but the cross-sections for the latter are about an order of magnitude smaller.
The gaugephilic/gaugephobic distinction does not make a qualitative difference.
Below the $\eta=0.5$ boundary, \rhod decays primarily to \pid pairs, and this is enhanced at the threshold.
Above, \rhod decays to pairs of fermions dominate. White areas indicate regions where the estimated cross-section is below $10^{-3}$ fb.}
\label{fig:bsm-production}
\end{figure}

Once produced, the decay of a \rhod depends on the mass hierarchy of the dark mesons. If $\rhod \rightarrow \pid\pid$ is kinematically allowed 
($\eta\leq 0.5$), then this is by far the dominant decay mechanism, with over 99\% branching fraction (and sub-percent level fractions of decays to 
fermions or quarks). On the other hand, if the \rhod cannot decay to \pid, it decays 25\% of the time to each generation of quark, 
with the remaining branching fraction is shared equally between decays to each generation of leptons.

Pair-production of \pid via a virtual \rhod is maximised when \mrhod is twice \mpid ($\eta = 0.5$), as expected. 
For the \sutl model, this process can reach about $10^4$~fb for $\mrhod\sim 1$~TeV, or $10^3$~fb in the \sutr model. 
Dark pions can also be singly-produced with a quark or gluon. 
This process is independent of the mass of the \rhod, and is typically of the order of 1-10~fb for a pion mass of 1~TeV.
Finally, $s$-channel production of pairs of leptons or quarks (and also $t$-channel for quarks) via a \rhod can be large, 
around 10~fb in the \sutr model and 10-100~fb for the \sutl model, where in addition lepton-neutrino pairs can be produced if the intermediate 
particle was a charged \rhod.
Examples of some of the subprocess cross-sections as a function of $\eta$ and \mpid for the processes mentioned above are shown in 
Fig.~\ref{fig:bsm-production}, for 13~TeV $pp$ collisions.

\subsection{Expected experimental signatures at the LHC}

Given the cross-section results, the LHC experimental signatures which are expected to be the most sensitive to these models will also 
depend on the mass hierarchy of the \rhod and \pid, as well as the handedness and the gaugephobic or gaugephilic nature of the model. 
For $\eta > 0.5$, the singly-produced \rhodz decaying to leptons can provide a clean signature that should be visible in high-mass 
DY measurements and searches. This is the case for all types of model considered.
One may also expect a large cross-section of dijet, or $t\bar{t}$-like signatures (from \rhod decays to quarks), 
but such hadronic-only signatures would be far more difficult to distinguish from the QCD background at the LHC.
For $\eta < 0.5$, the expected signatures depend more on the \pid decay modes: in gaugephobic models, the dominant decay of charged and neutral \pid 
is to a mixture of third-generation quarks (once above the kinematic threshold), leading to a multijet signature, suggesting that measurements 
of $t\bar{t}$, and/or final states involving $b$-tagged jets, will be most sensitive. 
Otherwise, one would need to rely on the accompanying decay of a vector boson to select events: 
this means that $Z+$jets or $W$+jets signatures might be expected to give good sensitivity. 
On the other hand, for gaugephilic models there are regions where the dominant branching fraction for \pid particles involves a vector boson 
and a Higgs boson, with decays to third generation quarks still important depending upon the \pid mass region. 
In this case, $Z+$jets or $W$+jets signatures can also be expected to play a role.

\section{Collider constraints on Dark Meson Production}
\label{sec:contur}

We scan over the parameter planes of the model, and use \CONTUR v2.1.1 to identify parameter points for which an observably 
significant number of events would have entered the fiducial phase space of the measurements available in \rivet 3.1.4~\footnote{With an additonal patch, now included in Rivet 3.1.5, to correct the normalisation of the ATLAS dilepton search.},.
\CONTUR evaluates the discrepancy this would have caused, under the assumption that the measured values, which have all been shown to
be consistent with the SM, are identical to it. This is used to derive an exclusion for each parameter
point, taking into account correlations between experimental uncertainties where available. The ATLAS run 2 dilepton resonance search~\cite{Aad:2019fac}
is also available in \rivet and is used, with some caveats since in this case the SM background is modelled by a fit 
to data rather than a precision calculation; the impact of this will be discussed below. 

\begin{figure}[h]
  \centering
  \subfloat[]{\includegraphics[width=0.32\textwidth]{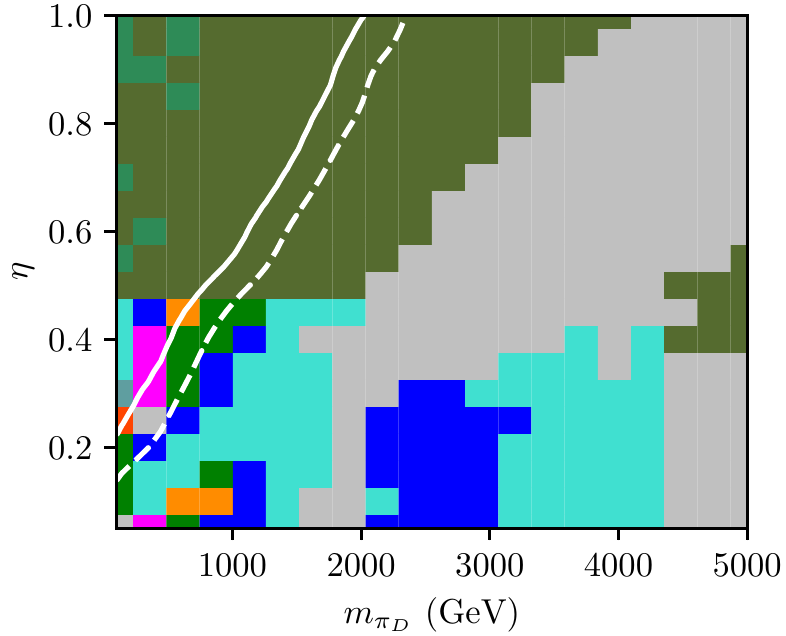}\label{fig:gil}}
  \subfloat[]{\includegraphics[width=0.32\textwidth]{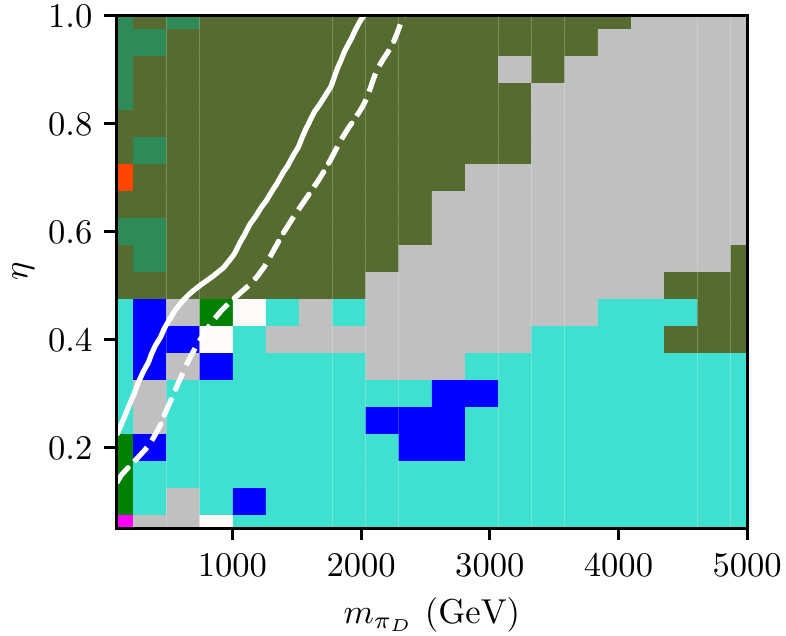}\label{fig:gol}}
  \subfloat[]{\includegraphics[width=0.32\textwidth]{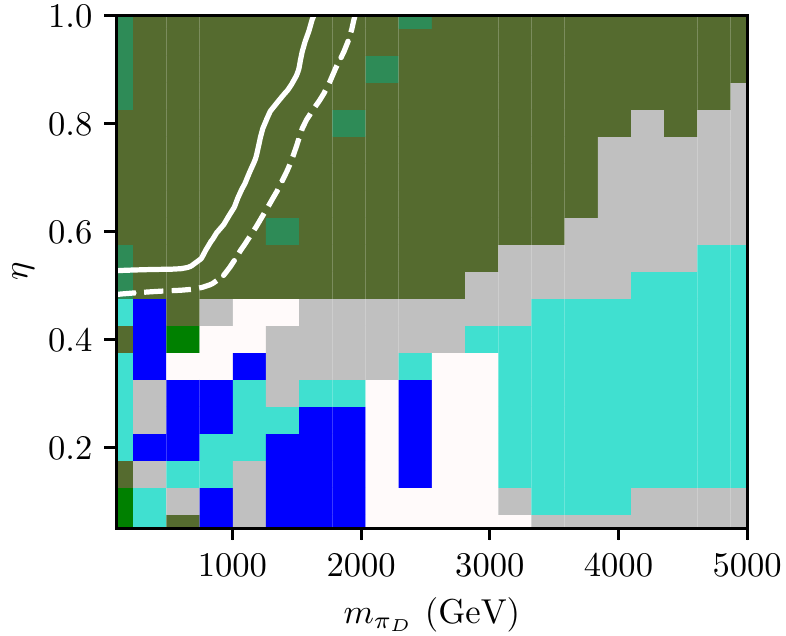}\label{fig:gor}}
  \caption{Scans in $\eta \-- \mpid$ for three sub-models.
  (a) Gaugephilic \sutl, (b) Gaugephobic \sutl (c) Gaugephobic \sutr. The colours
indicate the dominant signature pool giving the sensitivity. The white solid line is the 95\%
exclusion and the white dashed line is the 68\% exclusion.
  \label{fig:scans}}
    \begin{tabular}{llll}
        \swatch{seagreen}~CMS high-mass DY $\ell\ell$ &
        \swatch{darkolivegreen}~ATLAS high-mass DY $\ell\ell$ &
        \swatch{turquoise}~ATLAS $\ell_1\ell_2$+\met{}+jet \\
        \swatch{orangered}~ATLAS $ee$+jet &
        \swatch{green}~ATLAS \met{}+jet &
        \swatch{silver}~ATLAS jets \\
        \swatch{snow}~ATLAS Hadronic $t\bar{t}$ &
        \swatch{magenta}~ATLAS 4$\ell$ &
        \swatch{blue}~ATLAS $\ell$+\met{}+jet \\
        \swatch{darkorange}~ATLAS $\mu\mu$+jet &
\end{tabular}
\end{figure}

Scans are performed in the $\eta \-- \mpid$ plane for the Gaugephilic \sutl, 
Gaugephobic \sutl and Gaugephobic \sutr sub-models, and shown in Fig.~\ref{fig:scans}.

\begin{figure}[h]
  \centering
  \subfloat[]{\includegraphics[width=0.60\textwidth]{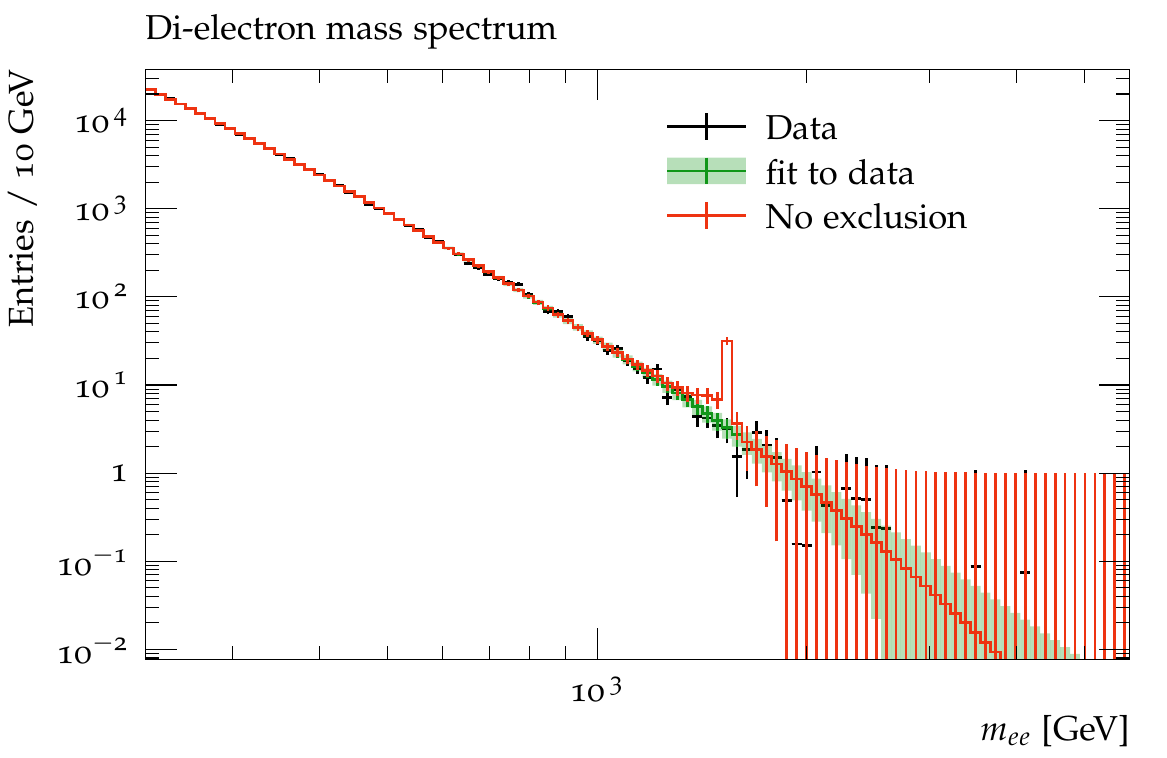}\label{fig:gol_hmdy}}\\
  \subfloat[]{\includegraphics[width=0.33\textwidth]{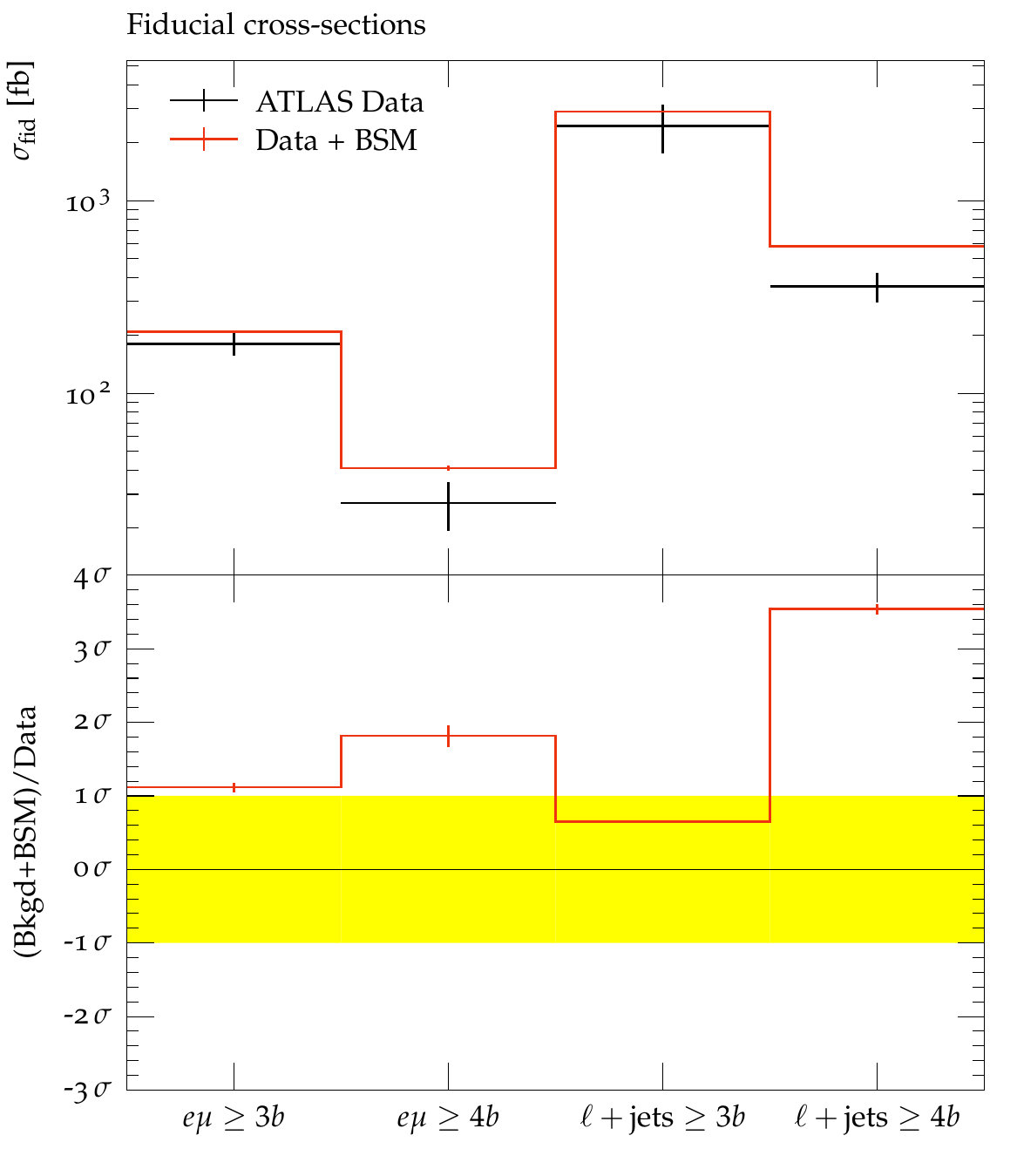}\label{fig:gol_ttbb}}
  \subfloat[]{\includegraphics[width=0.33\textwidth]{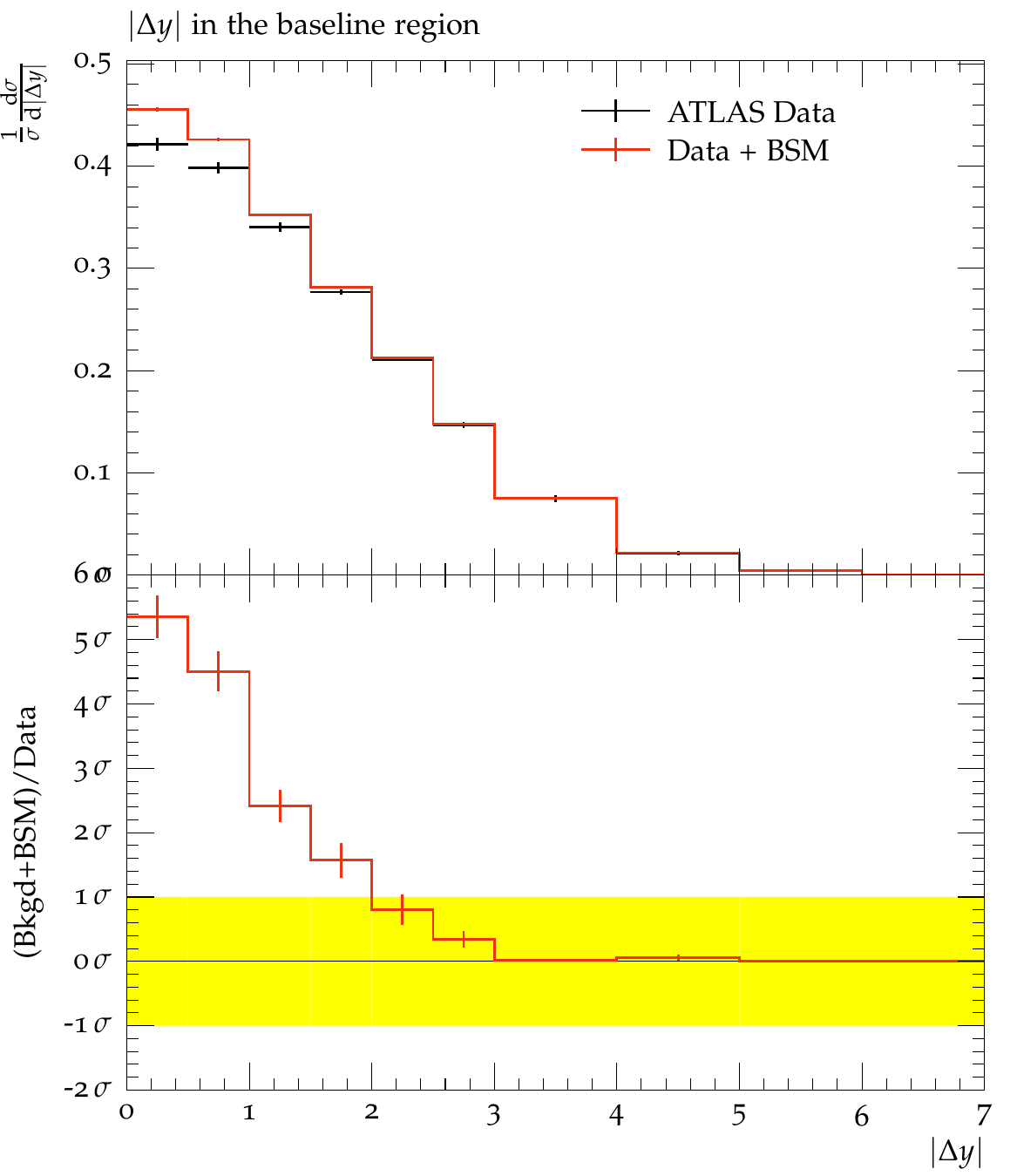}\label{fig:gol_lljet}}
  \caption{
  Example kinematic distributions for the Gaugephobic \sutl model.
  (a) Di-electron mass distribution from \cite{Aad:2019fac}, $\mpid=1.4$~TeV, $\eta=0.9$. 
  The uncertainties on the data and on the BSM 
  are statistical only, those on the background fit include the systematics.
  (b) $t\bar{t}b\bar{b}$ cross-sections~\cite{Aaboud:2018eki}, 
  $\mpid=390$~GeV, $\eta=0.45$. Full uncertainties are shown on the data, and by the
  yellow band in the lower ratio plot.
  (c) ATLAS dilepton+jets measurement~\cite{Aad:2014dta}, $\mpid=260$~GeV, $\eta=0.4$. 
  Uncertainties shown as in (c)
  \label{fig:rivet}}
\end{figure}

The expected change in behaviour around $\eta = 0.5$ is clearly seen. 
The dominant exclusion for much of the plane comes from the 139 fb$^{-1}$ ATLAS dilepton search, 
with the CMS measurement using 3.2 fb$^{-1}$~\cite{Sirunyan:2018owv} and the ATLAS 7 and 8~TeV measurements~\cite{Aad:2013iua,Aad:2016zzw} all
having an impact at lower \mpid.
No dilepton measurements using the full run 2 integrated luminosity from the LHC are
yet available.
For the $\eta > 0.5$ region, the \rhodz decay does indeed produce a resonant signature, as may be seen for an example point in Fig.\ref{fig:gol_hmdy}, 
and so this
limit can be taken as a good estimate. In this case, \rhod masses as high as 2~TeV are excluded for $\eta$ close to unity
in the left-handed cases, with this limit being reduced to around 1.7~TeV for the \sutr model due to the generally lower cross-section.
Example plots leading to the exclusion are shown in Fig.~\ref{fig:rivet}.

\begin{figure}[h]
  \centering
\subfloat[]{\includegraphics[]{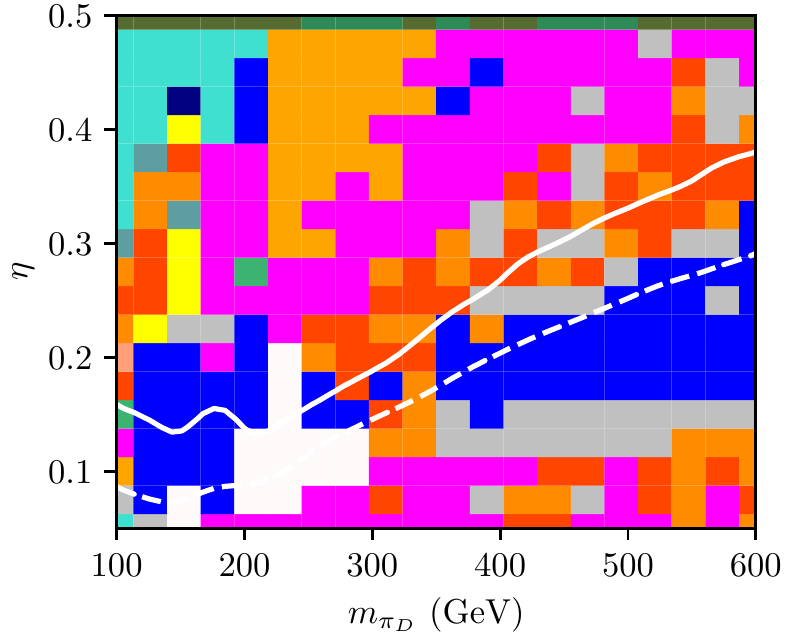}\label{fig:gil_loweta}}
\subfloat[]{\includegraphics[]{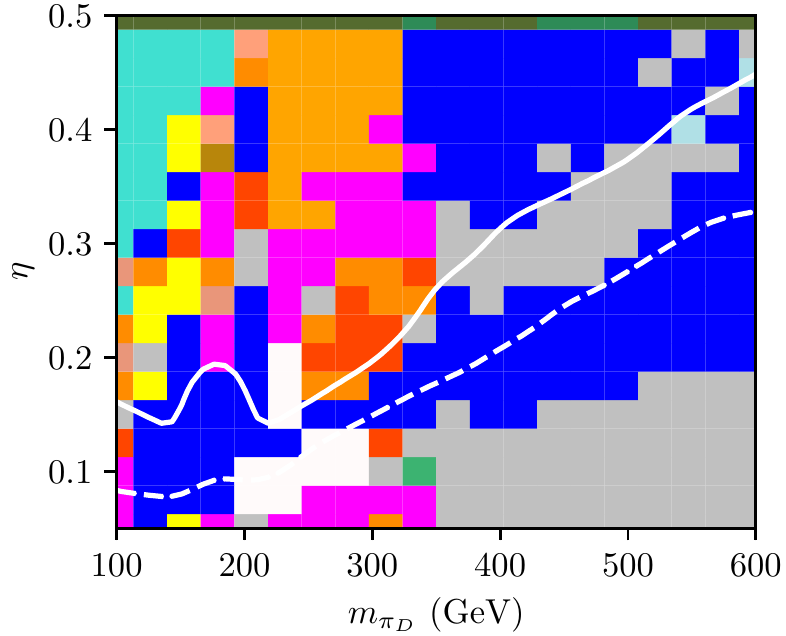}\label{fig:gol_loweta}}
  \caption{Scans in $\eta \-- \mpid$ for two sub-models, focussing on the $\eta<0.5$ region
  and excluding the ATLAS dilepton search.
  (a) Gaugephilic \sutl, (b) Gaugephobic \sutl 
The colours indicate the dominant signature pool giving the sensitivity. The white solid line is the 95\%
exclusion and the white dashed line is the 68\% exclusion.
  \label{fig:loweta_scans}}
    \begin{tabular}{llll}
        \swatch{lightsalmon}~CMS $ee$+jet &
        \swatch{darksalmon}~CMS $\mu\mu$+jet &
        \swatch{orangered}~ATLAS $ee$+jet \\
        \swatch{orange}~ATLAS $\ell\ell$+jet &                      
        \swatch{darkorange}~ATLAS $\mu\mu$+jet &

        \swatch{blue}~ATLAS $\ell$+\met{}+jet \\
        \swatch{navy}~ATLAS $\mu$+\met{}+jet &
        \swatch{cadetblue}~ATLAS $e$+\met{}+jet &
        \swatch{powderblue}~CMS $\ell$+\met{}+jet \\
        \swatch{snow}~ATLAS Hadronic $t\bar{t}$ &
        \swatch{turquoise}~ATLAS $\ell_1\ell_2$+\met{}+jet &
        
        \swatch{darkolivegreen}~ATLAS high-mass DY $\ell\ell$ \\
        \swatch{seagreen}~CMS high-mass DY $\ell\ell$ &        
        \swatch{magenta}~ATLAS 4$\ell$ &

        \swatch{yellow}~ATLAS $\gamma$ \\
        \swatch{mediumseagreen}~ATLAS $\ell\ell\gamma$ &
        \swatch{darkgoldenrod}~ATLAS $\gamma$+\met{} &
        
        \swatch{silver}~ATLAS jets 
    \end{tabular}
\end{figure}

This challenging $\eta < 0.5$ region requires further discussion, and in
Fig.~\ref{fig:scans} exclusions from a wide variety of other final states contribute,
reflecting the many possible \pid decay chains. 
For the \sutl cases, there is also some apparent sensitivity in this region from the dilepton search,
although it no longer domninates.
However, while lepton pairs may be produced in \pid decays, 
for example via top quarks, they are no longer resonant at the \rhod mass and so the ``bump hunt'' approach of \cite{Aad:2019fac} 
is no longer valid.
In this case, we revert to the default \CONTUR mode, which uses only particle-level measurements,
and re-scan the region $\eta<0.5$ at low \mpid.
This allows a more detailed investigation of 
the different \pid decay signatures. 
The results are shown in Fig.~\ref{fig:loweta_scans}.

With finer granularity of the scan, a mix of other analyses can now be seen to be contributing to the exclusion.
\begin{itemize}

\item when \mpid falls in the Higgs mass window of the ATLAS $H \rightarrow \gamma\gamma$ fiducial cross-section measurement~\cite{Aad:2014lwa},
the $\pid \rightarrow \gamma \gamma$ decays populate the cross-section and, despite their suppression due to dark flavour 
symmetry~\cite{Kribs:2018oad}, would have led to an observable excess at high $p_T^{\gamma\gamma}$, had they been present.

\item for $\eta \approx 0.2$ and $\mpid \approx 220$~GeV, the boosted hadronic top measurement~\cite{Aaboud:2018eqg} is most sensitive; 
tops are produced in the dominant decay modes of both neutral and charged \pid in this region, and for low $\eta$, 
where the \rhod is much heavier than the \pid, they will indeed be boosted.

\item at low \mpid and $\eta$ just below 0.5, the most sensitive measurements are of $t\bar{t}$ production in the $e\mu$ channel~\cite{Aad:2019hzw},
where an excess at low transverse momentum of the $e\mu$ system would have been particularly prominent.

\item at higher $\eta$ values, and especially in the gaugephilic case, the full run 2 four-lepton measurement~\cite{Aad:2021ebo} is sensitive 
for $\mpid \approx 200$~GeV and above. In this region $\pid\pid \rightarrow Z h t \bar{b}$ is significant, and can give rise to such
signatures, as can $\pid\pid \rightarrow hhZW$ and $\pid\pid \rightarrow hhWW$ (in the gaugephilic case) if kinematically possible.

\item at low \mpid and $\eta$, and in the gaugephobic case also at higher $\eta$ and $\mpid \approx 400$~GeV, 
lepton missing transverse energy and jet final states are important. In both cases this is due to 
$\pid\pid \rightarrow t\bar{t}t\bar{b}$, with the ATLAS $t\bar{t} + b$-jets measurement~\cite{Aaboud:2018eki} being especially powerful, 
where a significant excess in the $\ell + \geq 4b$ cross-section should have been visible, as shown in Fig.~\ref{fig:gol_ttbb}.

\item in the $0.3 < \eta < 0.5$ region for $\mpid > 220$~GeV, measurements of dilepton plus jets\cite{Aad:2014dta,Khachatryan:2016iob}
are sensitive, especially in the gaugephilic case. This is due the $\pid\pid \rightarrow Z h t \bar{b}$ 
(in both models) and the $\pid\pid \rightarrow hhZW$ and $\pid\pid \rightarrow hhWW$ decay modes (in the gaugephilic case). 
An example distribution is shown in Fig.~\ref{fig:gol_lljet}.

\item The high mass DY measurements~\cite{Aad:2016zzw,Sirunyan:2018owv} also still play a role in the total exclusion, although they
are rarely dominant. This indicates that future measurements should be powerful.

\end{itemize}

\section{Implications for Dark Matter Phenomenology}
\label{sec:dm}

Taking the searches considered in \cite{Kribs:2018ilo}, along with the constraints derived in the previous two
sections, we now consider the implications for stealth DM models. In Ref.~\cite{Appelquist:2015yfa}, the authors demonstrate potential ways to connect collider limits with DM analysis. 
We follow this strategy here in order to connect the different regimes together. 
It should be noted that unlike WIMP theories consisting of elementary particles, the LHC signals in our scenarios 
are not directly related to the production of dark matter. 
It is by using the underlying fundamental theory of dark quarks in the SU(4) fundamental representation, 
which fixes the mass spectra, that we can connect seemingly unrelated LHC signatures to DM phenomenology.

\begin{figure}[h]
  \centering
\includegraphics[scale = 0.5]{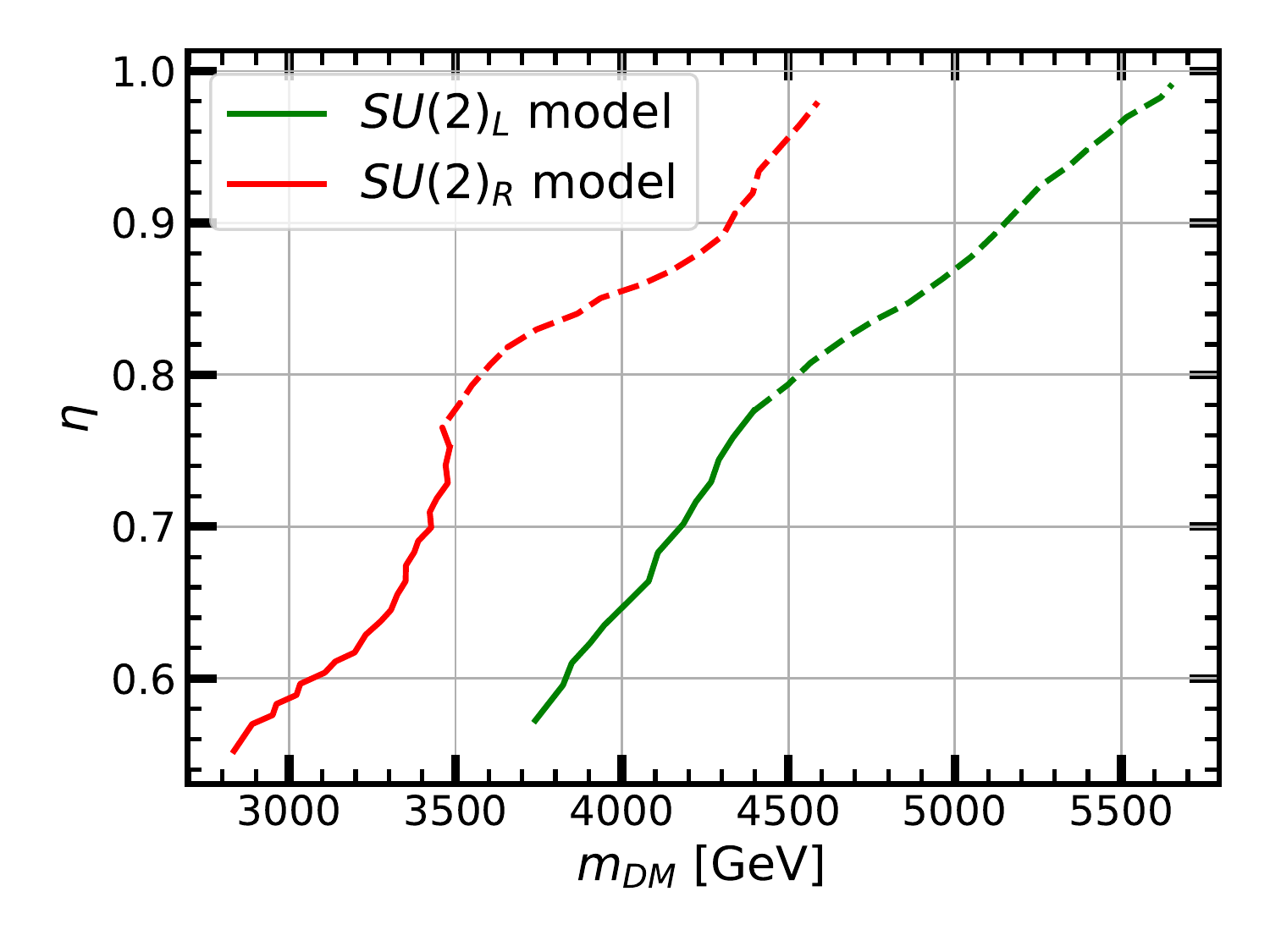}
  \caption{Limits on the mass of dark matter given constraints on \mpid and \mrhod for the \sutl model (green) and the \sutr model (red). 
As discussed in the text, only gaugephobic limits are used. The solid lines represent the region where lattice results 
are available, whilst the dashed lines represent our extrapolation assuming the heavy quark limit has been achieved. 
  \label{fig:DM}
  }
\end{figure}

From our results of the previous section, we obtain excluded values of $\eta$ and $\mpid$. 
These can be used to obtain the corresponding DM mass \mdm via
\begin{equation}
m_{S0}(\eta) = m_B(\eta) = \mdm(\eta) = \frac{amS0(\eta)}{amps(\eta)}\times \mpid(\eta)
\end{equation}
where $amS0, amps$ are the masses predicted by lattice simulations, and \mpid is the excluded pseudo-scalar mass obtained 
in the previous section. The exact values of $amS0, amps$ are summarised in Table~\ref{tab:lattice_inputs}.
We interpolate between different $\eta$ values in order to get Fig. \ref{fig:DM}. 
Finally, when $\eta \to 1$, we approach the heavy quark limit. In this limit, the masses of bound states are simply sum of the 
masses of the constituent quarks. 
In this case the scalar baryon mass, being made up of four quarks, should be twice \mpid, which has two constituent quarks. 
Having observed that such a limit is already reached for $\eta = 0.77$, we keep the mass ratios constant beyond for higher $\eta$ values. 

\begin{table}[h!]
\centering
\begin{tabular}{ |c|c|c|c|c|} 
 \hline
$\eta$ & $amps$ & $amv$ & $amS0$ & $f^{DM}_f$  \\ 
 \hline 
0.77 & 0.3477 & 0.4549 & 0.9828 & 0.153 \\
0.70 & 0.2886 & 0.4170 & 0.8831 & 0.262 \\
0.50 & 0.2066 & 0.3783 & 0.7687  & 0.338 \\
 \hline 
\end{tabular}
\caption{Lattice inputs for $\beta = 11.028$ on $32^3 \times 64$ lattices taken from~\cite{Appelquist:2014jch} for this work. $amps, amv$ and $amS0$ represent dimensionless pseudo-scalar, vector and dark baryon masses while $f^{(DM)}_f$ is lattice input for computing DM direct-detection cross-section via Higgs exchange.}
\label{tab:lattice_inputs}
\end{table}

The results are illustrated in Fig. \ref{fig:DM}, where we demonstrate exclusion limits on \mdm as a function of $\eta$. 
The two curves correspond to \sutl (green) and \sutr (red) gaugephobic models respectively. 
The dashed lines represent our extension of the provided lattice results to the heavy quark limits. 
We can see that the two limits differ significantly for smaller \mdm, where limits on $\eta$ in the \sutl theory are 
very different to those from \sutr theory, as discussed in the previous section. 
For higher \mdm the two scenarios lead to more similar limits on $\eta$, pushing the allowed \mdm to the multi-TeV regime. 
The limits from \sutr scenarios are always somewhat weaker than corresponding limits in \sutl theories.

\section{Combined Constraints}

As discussed previously, the DM candidate couples to the SM Higgs and scatters off nuclei at direct-detection experiments. 
The dark matter-nucleus scattering element constitutes of two parts, one corresponding to matrix element of the fermions in the scalar baryons ($f_f^{DM}$) which is computed via lattice simulation \cite{Appelquist:2014jch} and second, the Yukawa coupling between Higgs and dark quarks. Unfortunately, the Yukawa coupling as defined in Section~\ref{sec:model} and also introduced in Fig.~\ref{fig:model_details} can not be directly constrained and hence needs to be recast into so-called \yeff parameter which is independent of dark fermion mass \cite{Appelquist:2015yfa}. 
We identify that the so-called linear case as discussed in \cite{Appelquist:2015yfa} corresponds to the gaugephobic \sutl case, while the quadratic case 
corresponds to the \sutr gaugephobic case. This can be understood as follows. As explained in Section~\ref{sec:model}, the dark quarks obtain their masses from the vector as well as electroweak symmetry breaking (EWSB) mass terms. The physical masses are thus a mixture of the two mass terms after mass diagonalization. The coupling of Higgs to these physical quark masses will be proportional to the Higgs--dark quark Yukawa coupling times the mass of the dark quark, just like in the SM. If the dark quark masses are dominated by EWBS (chiral) mass term, one gets a quadratic dependence of Higgs--dark quark coupling, otherwise a linear dependence. Therefore, in the linear (\sutl) case, EWSB is dominantly (but not entirely) responsible for quark mass splitting, while the situation reverses in the other limit. It should be noted that \yeff is a free parameter of the theory and can be constrained using experimental data which is exactly what we will do in this section by combining latest Xenon1T limits with the LHC limits and obtain maximum allowed coupling \yeff between the dark quarks and Higgs. 
We briefly outline the exact procedure we follow in order to compute the DM--nucleus scattering cross-sections which closely follows\cite{Appelquist:2014jch, Appelquist:2015yfa}.

The dark matter nucleus spin-independent scattering cross-section is given by
\begin{equation}
\sigma(\textrm{DM},N) = \displaystyle\frac{\mu(m_{\textrm{DM}}, m_N)^2}{\pi} \left(Z \mathcal{M}_p + (A-Z) \mathcal{M}_n \right)^2,
\end{equation}
where $N$ is the nucleus, $\mu(m_{\textrm{DM}}, m_N)$ represents the nucleus - dark matter reduced mass\footnote{$\mu(m_1, m_2) = m_1 m_2/(m_1+m_2)$}, Z, A are the atomic and mass numbers, here taken for Xenon and finally $\mathcal{M}_{p,n}$ are the amplitudes for scattering off proton or neutron. 
This quantity is dependent on the experiment target, in order to ease the comparison among different experiments the cross-section is conveniently expressed as
\begin{equation}
\sigma_0(\textrm{DM}, a) = \sigma(\textrm{DM}, N)\displaystyle\frac{\mu(m_{\textrm{DM}}, m_a)^2}{\mu(m_{\textrm{DM}}, m_N)^2\,A^2},
\end{equation}
where $\sigma_0(\textrm{DM}, a)$ is the scattering cross-section per nucleon at zero momentum transfer.
The DM--nucleus scattering cross-section mediated by the Higgs exchange contains two parts, the first corresponding to the Higgs--SM quark current while the other contains 
Higgs--DM exchange. 

In particular the amplitude is given by\footnote{Note: we need to compute the squared amplitude, thu the expression for $\mathcal{M}_{p,n}$ must be squared.}

\begin{equation}
\mathcal{M}_{p,n} = \frac{g_{p,n}\,g_{DM}}{m_h^2}
\end{equation}
with 

\begin{equation}
g_{p,n} = \frac{m_{p,n}}{v}\left[\sum_{q = u,d,s}f_q^{(p,n)} + \frac{6}{27}\left( 1 - \sum_{q = u,d,s}f_q^{(p,n)} \right) \right] \\
\end{equation}
and
\begin{equation}
    g_{DM} \simeq
\begin{cases}
    \yeff\, f_f^{DM}, & \text{\sutl case}\\
    \\
    \yeff^2\,\displaystyle\frac{v}{\mdm} f_f^{DM},  & \text{\sutr case}
\end{cases}
\label{eq:yeff}
\end{equation}

\begin{figure}[h]
  \centering
  \subfloat[]{\includegraphics[scale = 0.5]{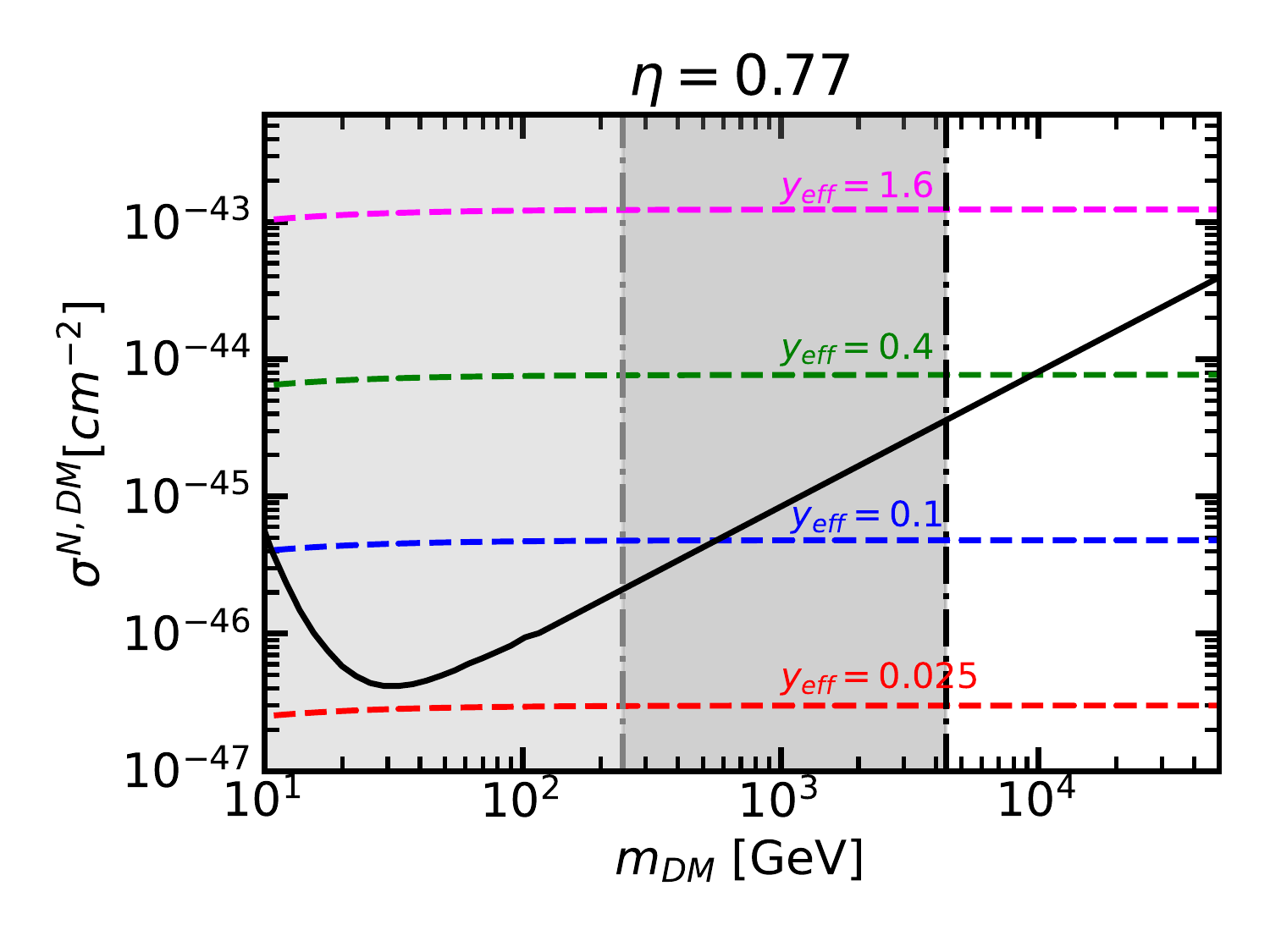}}
  \subfloat[]{\includegraphics[scale = 0.5]{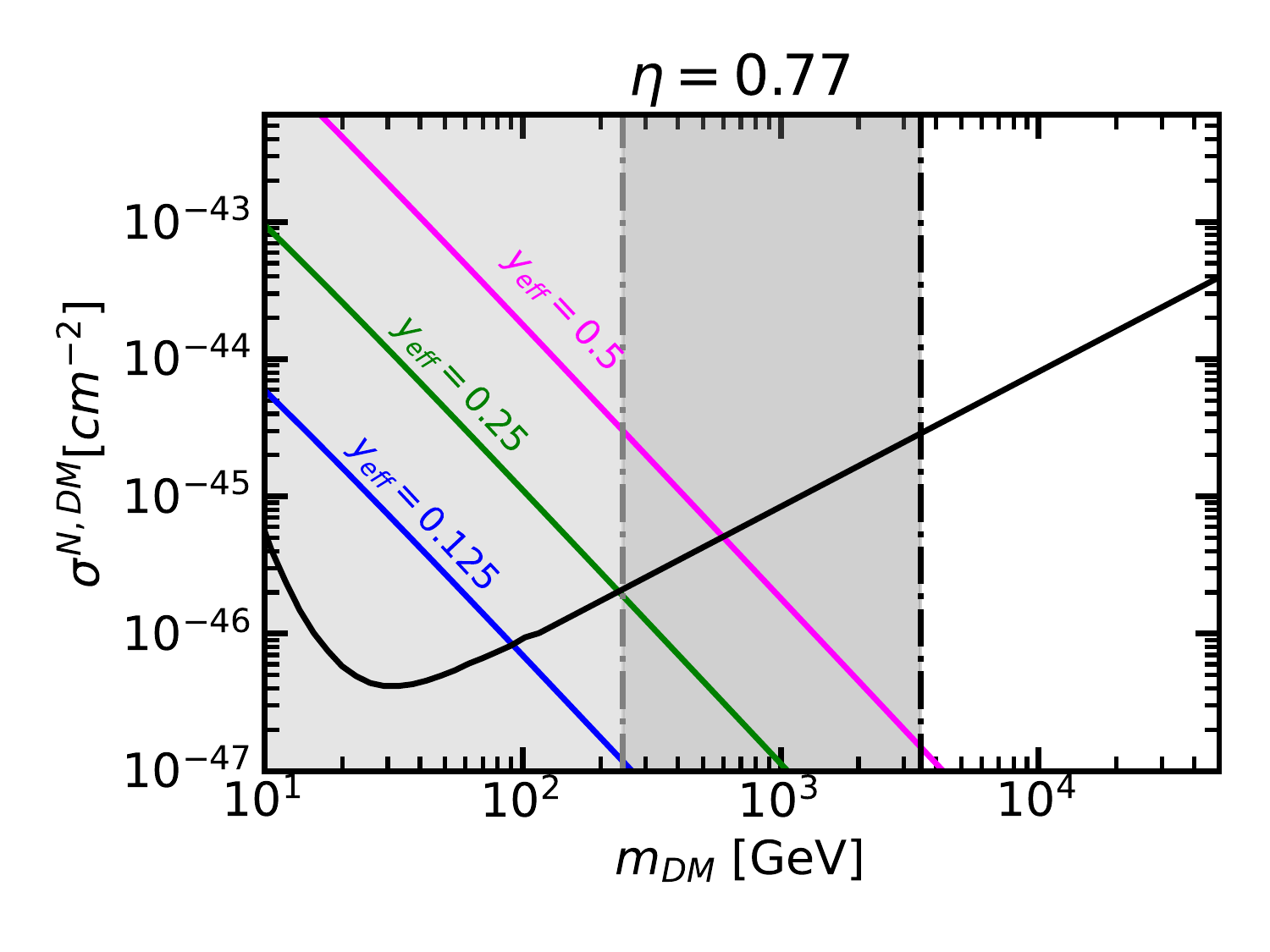}}\\
    \subfloat[]{\includegraphics[scale = 0.5]{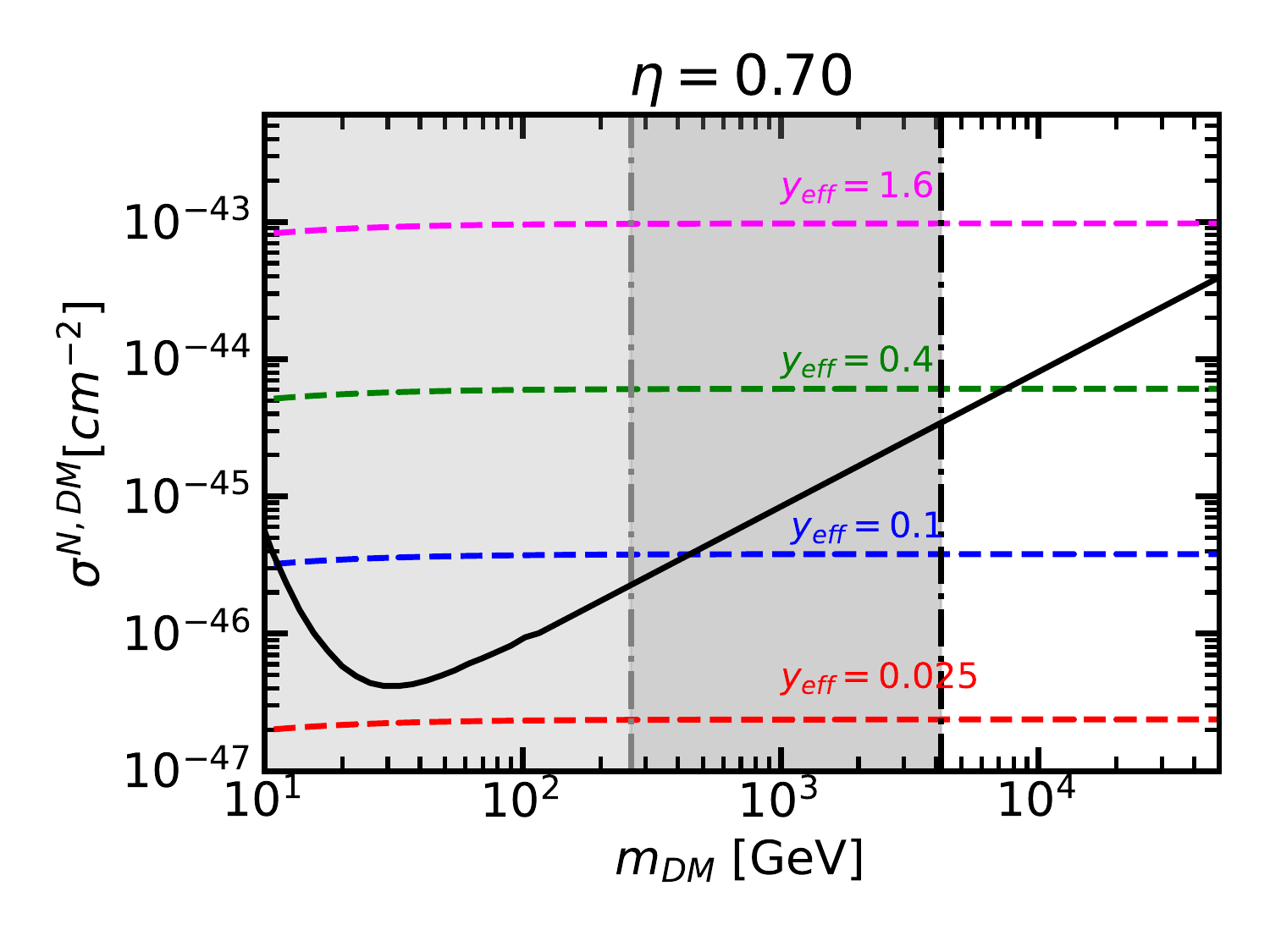}}
  \subfloat[]{\includegraphics[scale = 0.5]{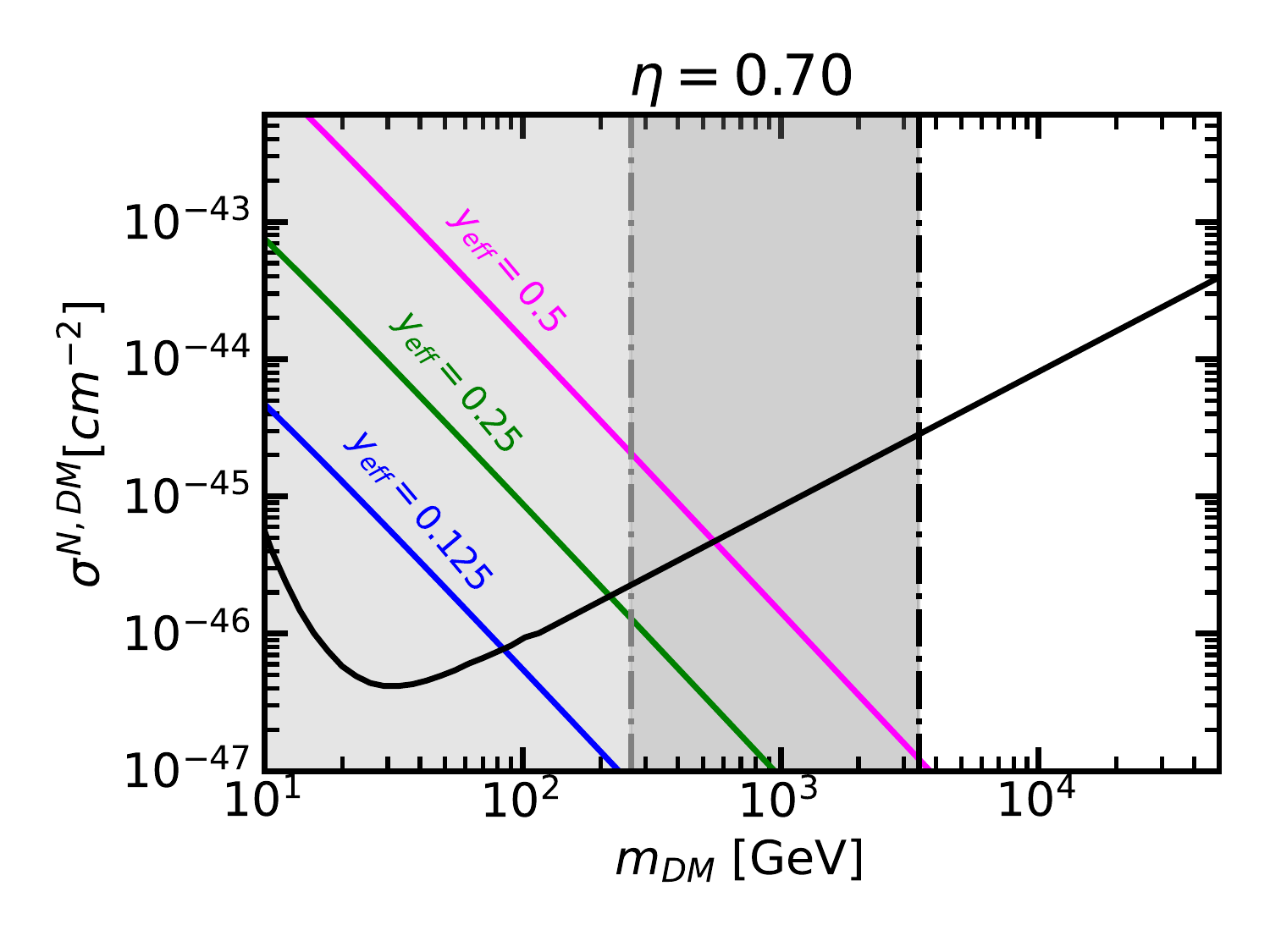}}\\
    \subfloat[]{\includegraphics[scale = 0.5]{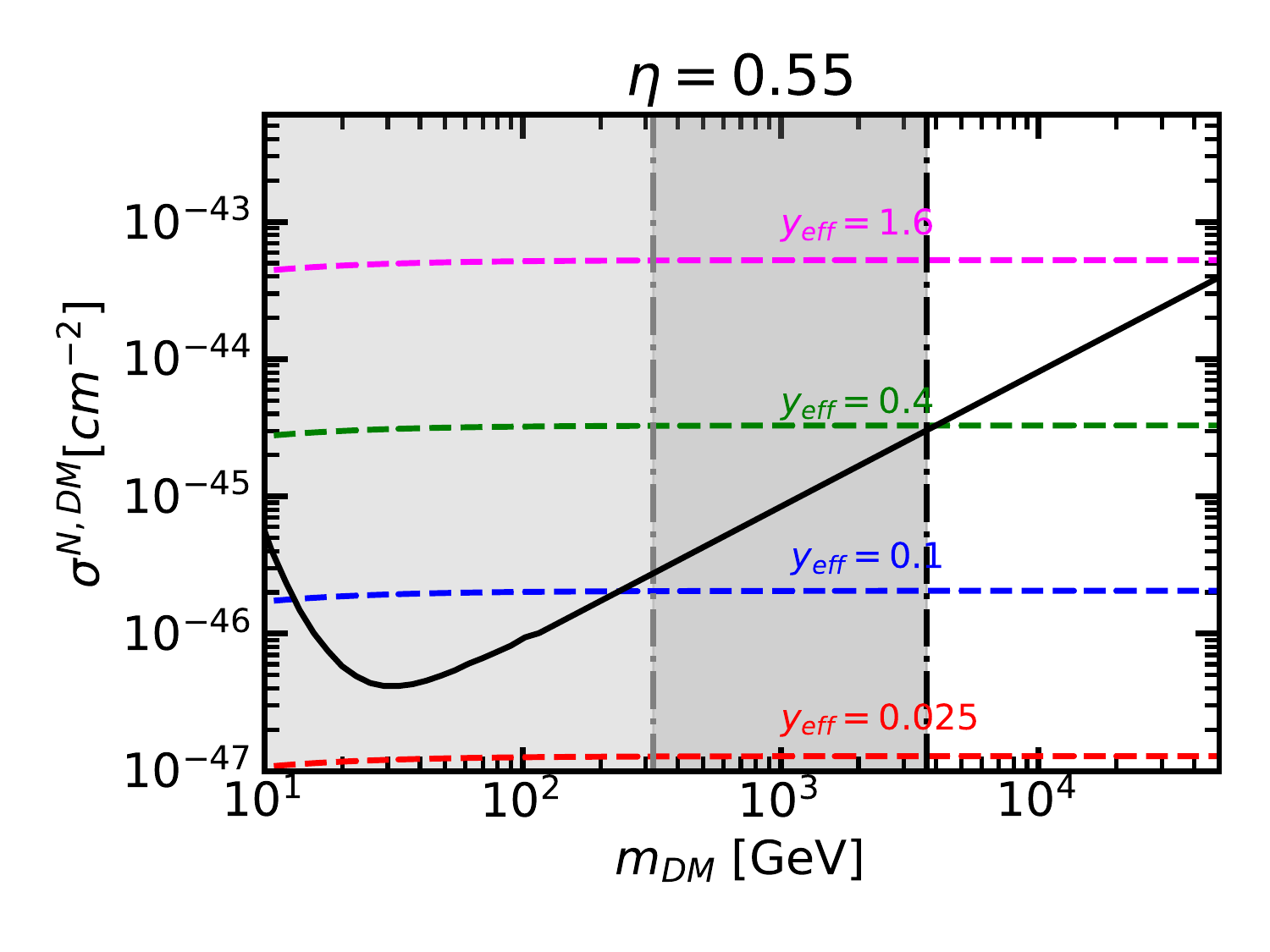}}
  \subfloat[]{\includegraphics[scale = 0.5]{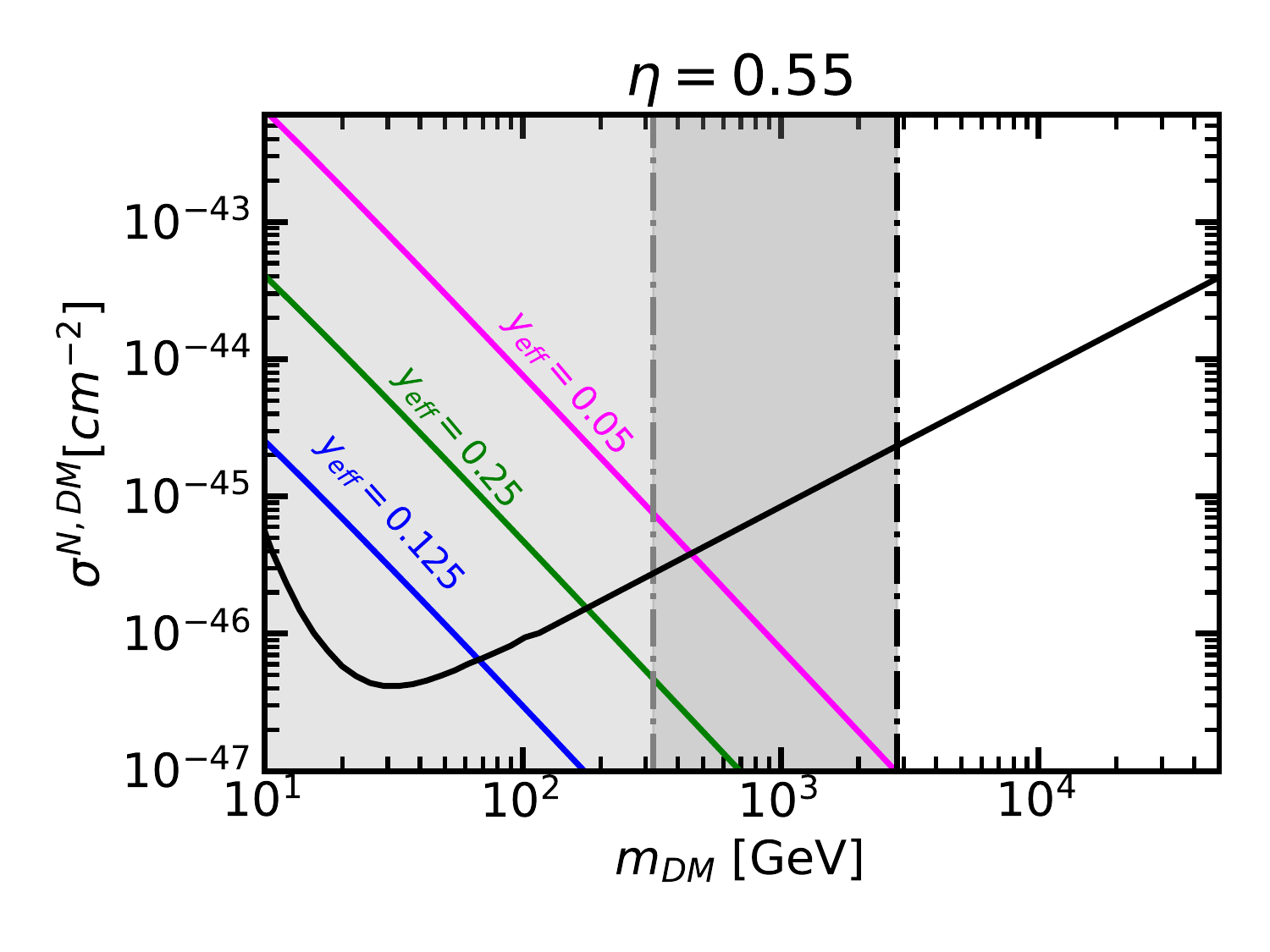}}
  \caption{Left panels (a,c,e), constraints for \sutr model and right panels (b,d,f), constraints $SU(2)_L$ model, for several values of
  $\eta$. Colored contours show DM--SM scattering cross-sections for fixed value of \yeff, vertical lines represent \mdm limits derived from LEP limits on pion mass (grey dot-dashed line) and updated LHC constraints derived in this work (black dot-dashed line). Also overlaid are the recent Xenon1T constraints on DM--nucleon coherent scattering. \cite{Aprile:2018dbl}.
  \label{fig:DM_combined}
  }
\end{figure}

\begin{figure}[h]
  \centering
  \subfloat[]{\includegraphics[scale = 0.5]{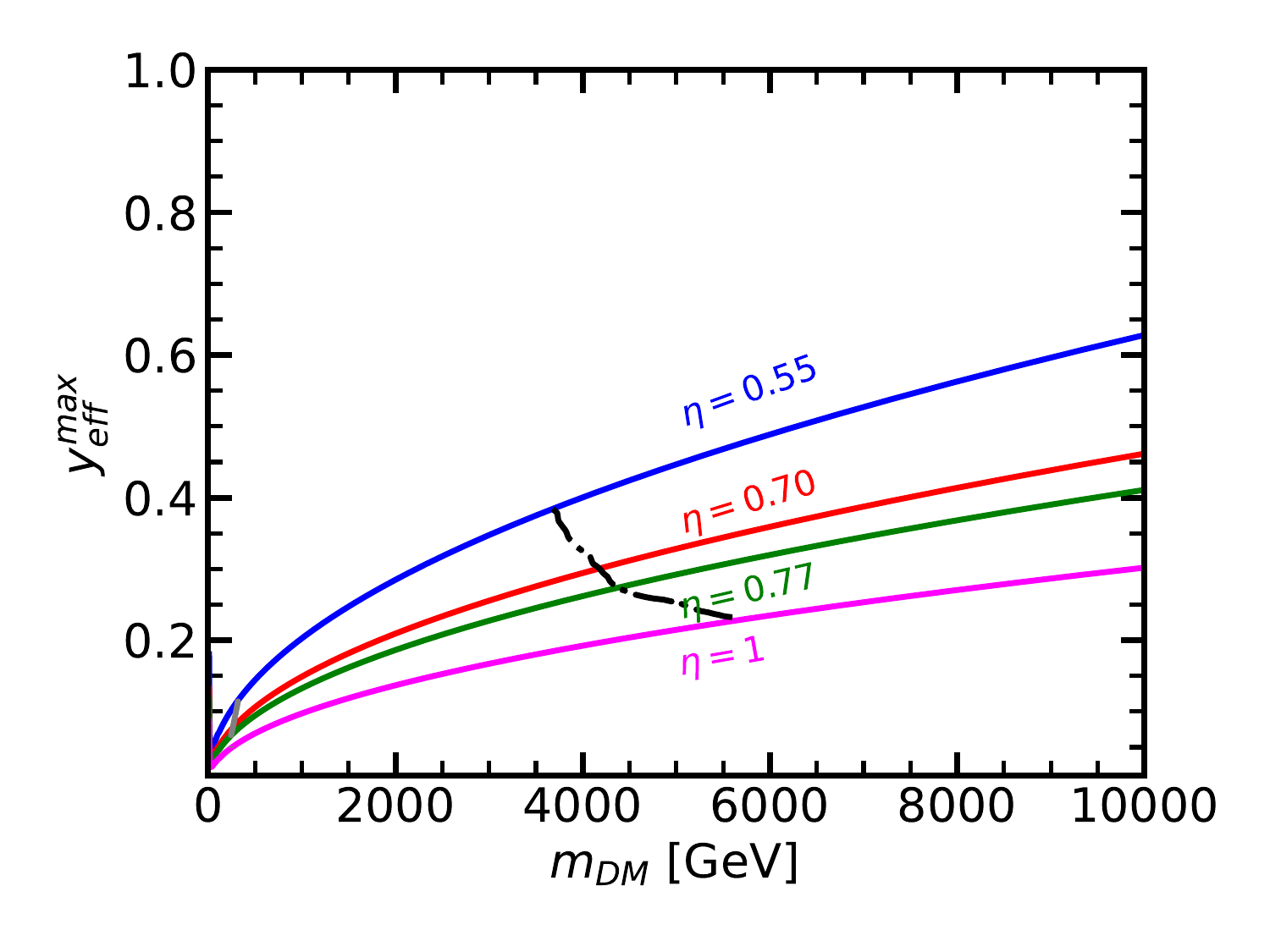}}
  \subfloat[]{\includegraphics[scale = 0.5]{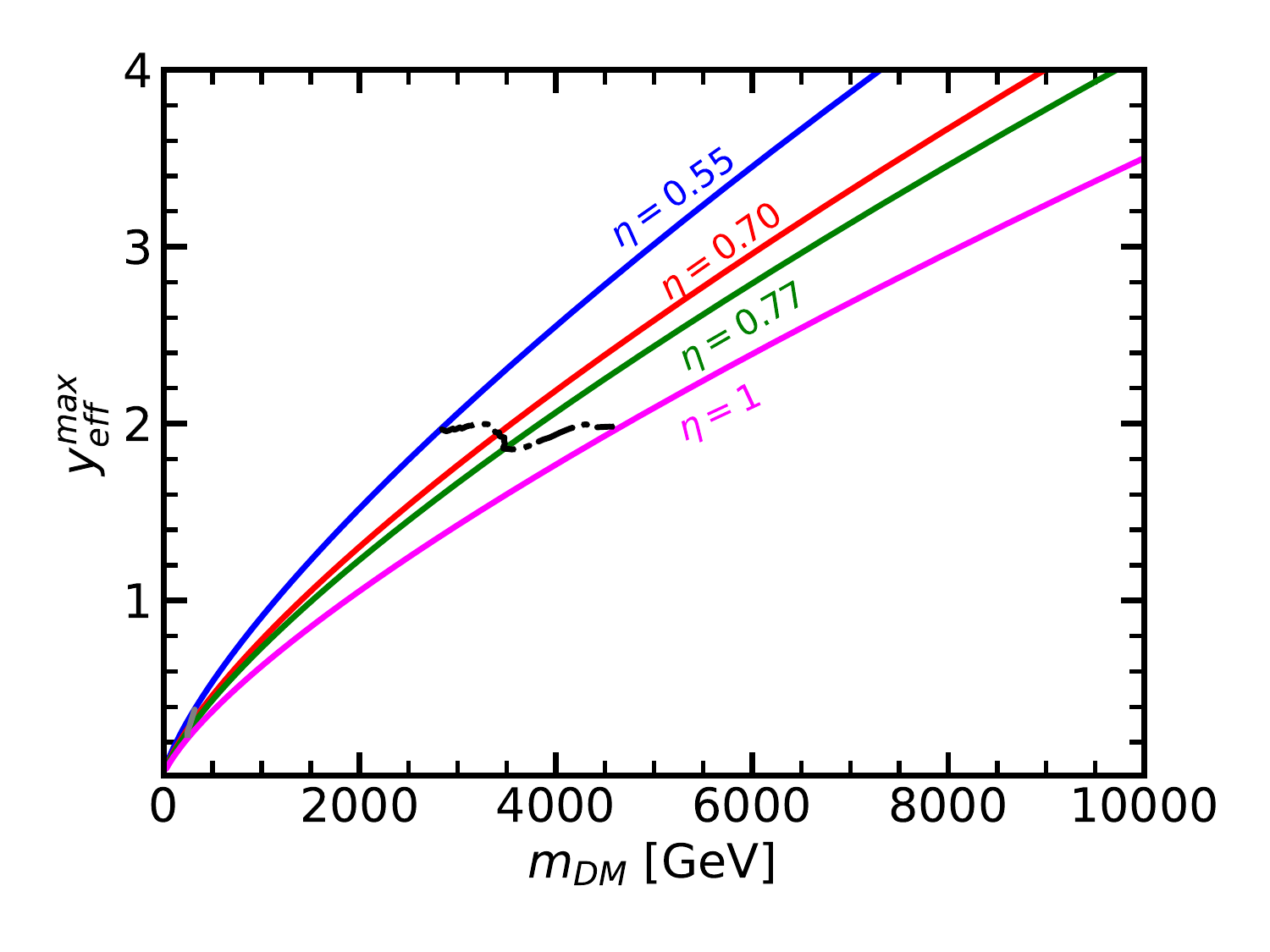}}
  \caption{(a) Maximum allowed value of \yeff for given value of $\eta$ for \sutl model, (b) constraints \sutr model. The black dot dashed line shows LHC limits derived in this work. Unlike Applequist et al\cite{Appelquist:2015yfa}, we do not extend our LHC limits in the chiral limit  (ie low $\eta$), meaning we have an abrupt cut-off in the LHC limits around \mdm of about 3.5 TeV (a) and 3 TeV (b) respectively. 
  \label{fig:DM_combined_yeff}
  }
\end{figure}

The quantity $f_f^{DM}$ is extracted from lattice calculations and precise values are tabulated in Table~\ref{tab:lattice_inputs}, while \yeff is the effective Yukawa coupling independent of the dark quark masses.  Using the definitions above and the lattice parameters for $f_q^{(p,n)}$ as defined in \cite{Hill:2011be, Belanger:2008sj}, we update the constraints on \yeff, as shown 
in Fig.\ref{fig:DM_combined} and Fig.\ref{fig:DM_combined_yeff}\footnote{We thank O.Witzel, E. Neil and G. Kribs for confirming that the original constraints for quadratic case correspond to contours of $\yeff^2$ instead of $\yeff$.}.

To begin with, in Fig.\ref{fig:DM_combined}, we show contours of \yeff in the  \mdm and DM--nucleon scattering cross-section plane for \sutr model (left panels) and \sutl case (right panels) for fixed value of $\eta$ of 0.77 (a,b), 0.70 (c,d) and 0.55 (e,f). The values of $\eta$ correspond to those for which lattice calculations are performed. We also overlay the latest Xenon1T DM--nucleon elastic scattering cross-section limits~\cite{Aprile:2018dbl}. The way to interpret this plot would be to find values of \yeff  which lie below the Xenon1T curve for a given \mdm. These correspond to direct detection allowed parameter spaces. Finally, we also show illustrate exclusions on \mdm obtained from LEP II limits~\footnote{Although we note some caveats to the oft-quoted 
generic 100~GeV limit on charged fermions from LEP-II~\cite{Egana-Ugrinovic:2018roi}, they do not apply here.} in~\cite{Kribs:2018oad} (dot-dashed grey line) and from our LHC analysis as discussed in section above (dot-dashed black line). Combining this with the DM--nucleon scattering cross-section limits, one can then derive the maximum value of \yeff which obeys all limits. Stringent limits on \yeff imply weaker coupling between DM and Higgs and thus in general weaker DM--SM interactions. 

Before discussing the ultimate lessons we learn from this, we first discuss Fig.\ref{fig:DM_combined}. As expected from Eq.\ref{eq:yeff}, for the \sutr case the DM--nucleon scattering cross-section is independent of \mdm while for \sutl case it is inversely proportional to \mdm for a given value of \yeff.  For \sutr case, the Xenon1T allowed value of \yeff decreases as $\eta$ increases, this is because $f^{DM}_f$ increases as $\eta$ increases. The dependence is more complicated for \sutl case where $g_{DM}$ depends on both $f^{DM}_f$ and \mdm which increase as $\eta$ increases. However $f^{DM}_f$ increases faster than \mdm and hence \yeff decreases albeit at a smaller rate.

Finally, the limits from DM--nucleon elastic scattering are complemented by the limits on \mdm derived by LEP and LHC analyses. In particular, for each $\eta$, the corresponding figure illustrates the impact of updated LHC limits compared to the LEP exclusion. The LHC limits as obtained by us in this analysis together with the Xenon1T limits thus either demand smaller values of \yeff or large \mdm to be compatible with current experimental situation. This is precisely reflected in Fig.~\ref{fig:DM_combined_yeff}, where we plot the maximum allowed values of \yeff for various value of $\eta$ (solid lines) and overlay the limits obtained from LHC (dot-dashed line). It should be noted that lower values of $\eta$ push the theory into the chiral (massless quark) limit where no lattice results are available, and we refrain from deriving any limits in this region.

\section{Relic Density Implications}

Global investigations, first carried out in Ref.~\cite{Kribs:2018oad, Appelquist:2015yfa, Kribs:2018ilo} and this analysis, can have implications 
for possible relic density generation mechanisms in the underlying composite SU(4) theories.
As argued in Ref.~\cite{Appelquist:2015yfa}, there are two broad avenues to obtain relic density.
The first is so-called symmetric abundance mechanism, where annihilation of dark baryons into dark mesons controls the relic density
in a way akin to the WIMP mechanism. Being entirely within the dark sector, this annihilation channel is expected to dominate.
From simple dimensional analysis, one expects baryon masses to be $\sim 100\,\rm{TeV}$ for relic abundance to be satisfied.
However, a precise calculation of this requires knowledge of annihilation channels. Within the framework
of chiral perturbation theory, this means one needs to include baryons, which is non-trivial.
The other way is to perform lattice simulations of scattering amplitudes, which are absent for the model we work with.

The other possibility is to obtain relic abundance from the decays of the electroweak sphaleron -- the non-perturbative solution at finite
temperature that allows transition between different vacua with different $B + L$ numbers.
This corresponds to the so-called asymmetric mechanism. Given that dark fermions are charged under the SM gauge group,
the electroweak sphalerons also contribute to generation of dark baryons during the decay after sphaleron ``freeze-out''.
Whether such decays generate any asymmetry, which further affects dark baryon number generation is also an important consideration.
In cases where there is no asymmetry, previous calculations for techibaryon models~\cite{Nussinov:1985xr, Chivukula:1989qb, Barr:1990ca}
have shown viable dark baryon masses in 1-2 TeV ranges. However these technibaryon models rely on dark fermion mass generation purely
from the EWSB mechanism. In this work, as the dark fermion masses are a combination of vector confinement and the EWSB mechanism,
one can qualitatively understand the relic density generation mechanisms, but quantitative estimates are harder.
If there is an initial asymmetry, e.g.~in a globally conserved quantity, one could also obtain the correct abundance.

Given the model dependence of these calculations, and absence of lattice results, we refrain from making further statements about dark baryon
relic density. However, we note that the LHC limits presented here and in Ref.~\cite{Kribs:2018oad, Appelquist:2015yfa, Kribs:2018ilo}
can be relevant for scenarios in which the DM relic density is generated via sphaleron processes.
Finally, we also note that relic density generation via symmetric mechanisms (i.e.~when annihilation proceeds within dark sectors)
will still be a viable avenue.

\section{Conclusions}

The analysis presented in this paper is significant in two respects. First, we take into account the updated Xenon1T limits which are
almost an order of magnitude stronger than the previous iteration. Second, the updated LHC limits (now including those from measurements
as well as searches) push \mdm to the multi-TeV regime, and much stronger than any previously obtained. 

When the \rhod decays directly to SM particles, these limits probe some scenarios in which the DM relic density is generated via
sphaleron processes. However, the LHC limits as derived in~\cite{Kribs:2018ilo} and in this work are relatively weak for small $\eta$,
where $\rhod \to \pid\pid$ decays are open. 
Small values of $\eta$ also correspond to the chiral limit of the theory. Therefore we conclude that on the theory side,
more efforts are needed to constrain chiral limit of such strongly-interacting scenarios.
From the experimental point of view, limits on this region can be expected to improve as more precise measurements of a variety
of final states are made, including tops, $b$-quarks and dileptons.
As pointed out in~\cite{Kribs:2018ilo}, for $\mpid < 150$~GeV or so, final states involving $\tau$ leptons are important.
No such measurements are currently available; should they be made in future, they could have significant impact.

Overall, while some the general properties of these models can be surmised, the detailed phenomenology has many uncertainties, in particular
driven by the unknown strong dynamics.
This makes dedicated search analyses for specific benchmark scenarios of limited interest.
While scenarios (as mentioned in the introduction) which lead to non-SM like final states will still require dedicated search strategies,
the uncertainty in the phenomenology strongly motivates model-independent measurements of a wide variety
of SM-like final states. These measurements can be then be analysed for discrepancies and, in their absence, rapidly interpretated as limits
over a very broad range of parameters. 

\section*{Acknowledgements}
SK thanks Axel Maas for several useful discussions and E. Neil, O. Witzel and G. Kribs  for clarifications about the direct-detection constraints. We thank Chi Leung for precursor investigations
of the model as part of her final year undergraduate project.
JMB has received funding from the European Union's Horizon 2020 research and innovation
programme as part of the Marie Skłodowska-Curie Innovative Training
Network MCnetITN3 (grant agreement no. 722104) and from a UKRI Science and Technology Facilities Council
(STFC) consolidated grant for experimental particle physics. SK is supported by Elise-Richter grant project number V592-N27.

\appendix

\FloatBarrier
\section{Collider exclusions for the $\nd=2$ case}
\label{app:ndark2}

\begin{figure}[h]
  \centering
  \subfloat[]{\includegraphics[width=0.32\textwidth]{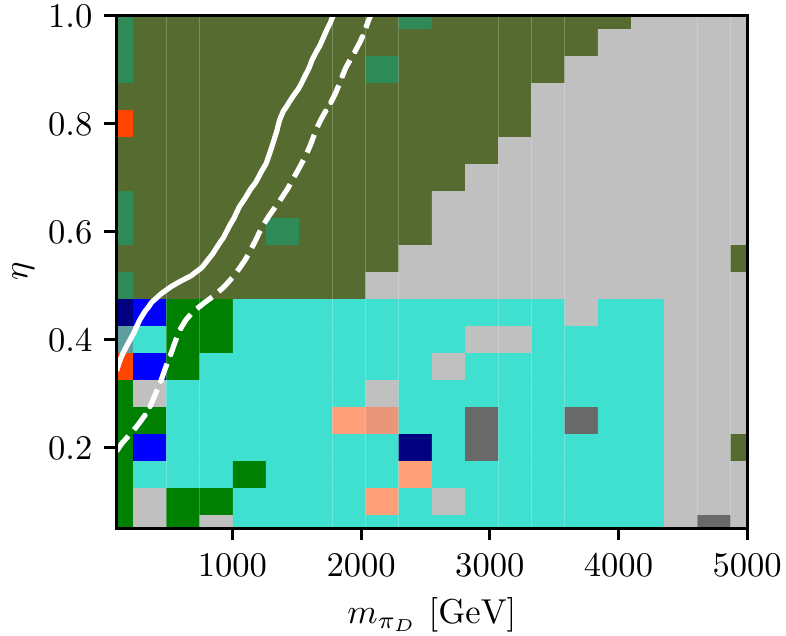}\label{fig:gil_nd2}}
  \subfloat[]{\includegraphics[width=0.32\textwidth]{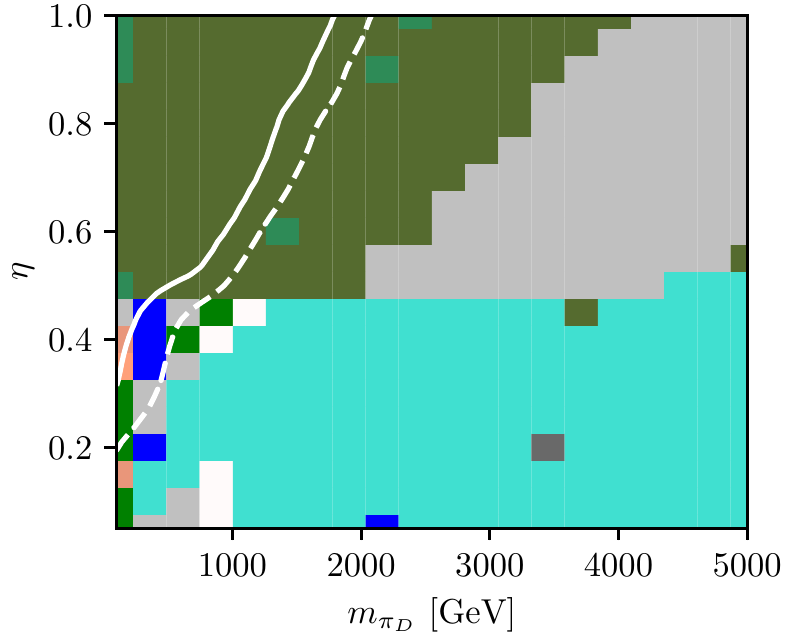}\label{fig:gol_nd2}}
  \subfloat[]{\includegraphics[width=0.32\textwidth]{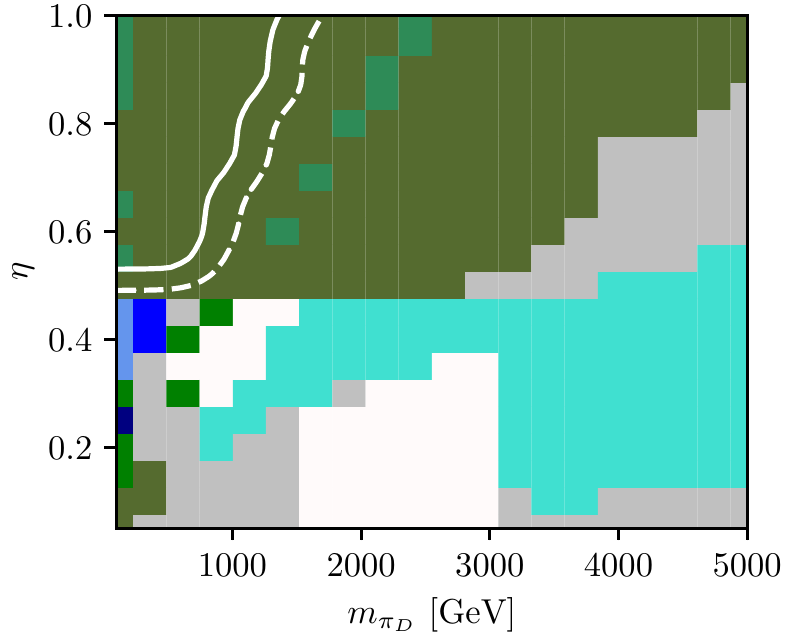}\label{fig:go_nd2r}}

  \caption{Scans in $\eta \-- \mpid$ for three sub-models.
  (a) Gaugephilic \sutl, (b) Gaugephobic \sutl (c) Gaugephobic \sutr. The colours
indicate the dominant signature pool giving the sensitivity. The white solid line is the 95\%
exclusion and the white dashed line is the 68\% exclusion.
  \label{fig:scans_nd2}}
    \begin{tabular}{llll}
        \swatch{blue}~ATLAS $\ell$+\met{}+jet &
        \swatch{cadetblue}~ATLAS $e$+\met{}+jet &
        \swatch{cornflowerblue}~ATLAS $\ell_1\ell_2$+\met{} \\
        \swatch{darkgoldenrod}~ATLAS $\gamma$+\met{} &
        \swatch{darkolivegreen}~ATLAS high-mass DY $\ell\ell$ &
        \swatch{darkorange}~ATLAS $\mu\mu$+jet \\
        \swatch{deepskyblue}~CMS $e$+\met{}+jet &
        \swatch{navy}~ATLAS $\mu$+\met{}+jet &
        \swatch{green}~ATLAS \met{}+jet \\
        \swatch{magenta}~ATLAS 4$\ell$ &
        \swatch{orangered}~ATLAS $ee$+jet &
        \swatch{seagreen}~CMS high-mass DY $\ell\ell$ \\
        \swatch{silver}~ATLAS jets &
        \swatch{snow}~ATLAS Hadronic $t\bar{t}$ &
        \swatch{turquoise}~ATLAS $\ell_1\ell_2$+\met{}+jet \\
    \end{tabular}
\end{figure}

All the results presented in the main body of the text assumed that $\nd$ parameter of the models under consideration was set to $4$. 
However, the sensitivity of LHC constraints vary considerably with this $\nd$ parameter.
To illustrate this effect, in this section we consider the $\nd=2$ case.
The phenomenology and dominant production and decay modes in the $\nd=2$ and $\nd=4$ cases are broadly comparable, but the overall cross-sections are reduced by a factor of approximately 2.
Analogously to the study presented in Section~\ref{sec:contur}, the $\mpid$ vs $\eta$ parameter planes of the Gaugephilic \sutl, Gaugephobic \sutl Gaugephobic \sutr we scanned using \CONTUR, but this time choosing $\nd=2$. The results are shown in Fig.~\ref{fig:scans_nd2}.

The resulting constraints have a similar structure to the $\nd=4$ case (which were shown in Fig.~\ref{fig:scans}).
In particular, for the left-handed models, the constraints cut across the \mpid vs $\eta$ plane, while for the right-handed model,
the sensitivity ends roughly below $\eta \sim 0.5$, as is expected.  

In this case, the dominant exclusion for much of the plane still comes from the 139 fb$^{-1}$ ATLAS dilepton search, 
with the CMS measurement using 3.2 fb$^{-1}$~\cite{Sirunyan:2018owv} and the ATLAS 7 and 8~TeV measurements~\cite{Aad:2013iua,Aad:2016zzw} 
having an impact at lower \mpid (although less so than in the  $\nd=4$ case).

However, the constraints are overall weaker than in the $\nd=4$ case, retreating in general by about $250$~GeV in \mpid. 
In the \sutr plane, the $\nd=2$ results also do not reach into the $\eta<0.5$ region as they did for $\nd=4$.

In summary, scanning the \sutl and \sutr models where the $\nd$ parameter is set to $2$ instead of $4$ leads to a very similar set of results, but weaker due to an overall lower cross-section for the key processes expected at the LHC. This illustrates that the results of the study presented in this paper are sensitive to choices in parameters such as $\nd$, and this should be kept in mind when interpreting these results.

\bibliography{hdm.bib,contur-anas.bib}


\end{document}